\documentclass[10pt,journal,compsoc]{IEEEtran}
\newif\ifpeerreview

\peerreviewfalse

\usepackage[nocompress]{cite}
\usepackage{url}
\usepackage{amsmath,amssymb,graphicx}

\usepackage{lipsum} 

\usepackage{subfigure}
\usepackage{graphicx}
\renewcommand\figurename{Figure} 
\usepackage{mathtools}
\usepackage{algorithm}
\usepackage{algorithmic}
\usepackage{amsmath}

\usepackage{booktabs}
\usepackage{xcolor}
\usepackage{stackengine}

\usepackage[switch]{lineno}

\title{Intelligent Autofocus}

\author{Chengyu Wang, Qian Huang, Ming Cheng, Zhan Ma, David Brady
\IEEEcompsocitemizethanks{\IEEEcompsocthanksitem Chengyu Wang, Qian Huang and David Brady are with the Department
of Electrical and Computer Engineering, Duke University, Durham,
NC, 27708.\protect\\
E-mail: david.brady@duke.edu
\IEEEcompsocthanksitem Ming Cheng and Zhan Ma are with School of Electronic Science and Engineering, Nanjing University, Nanjing,
Jiangsu, China, 210023}
}

\begin{document}

\IEEEtitleabstractindextext{%
\begin{abstract}
We demonstrate that deep learning methods can determine the best focus position from 1-2 image samples, enabling 5-10x faster focus than traditional search-based methods. In contrast with phase detection methods, deep autofocus does not require specialized hardware. In further constrast with conventional methods, which assume a static "best focus," AI methods can generate scene-based  focus trajectories that optimize synthesized image quality for dynamic and three dimensional scenes. 

\end{abstract}

}

\maketitle

\IEEEraisesectionheading{
  \section{Introduction}\label{sec:introduction}
}
%
%
%
%
\IEEEPARstart{D}{espite} continuing advances in digital and computational imaging, mechanical functions, e.g. focus and zoom, remain central to camera operations. As discussed in~\cite{parallelCam}, it is very challenging for computation to recover the clarity of information captured by a well-focused image. For this reason, the science and art of autofocus (AF) is well developed in camera systems. To date AF has relied on three strategies: (1) active range sensing, (2) phase detection and (3) contrast maximization. Active sensing uses structured illumination, time of flight sensing, or ultrasonic sensing to sense the range to objects of interest and adjust the focus accordingly. Phase detection uses light field sensors to measure the disparity between the current focus setting and the in focus setting. Contrast maximization uses image quality measures to evaluate focal quality and search for optimal focus. Each strategy has advantages and disadvantages. Active illumination and phase detection can determine the correct focus in a single time step, and thus can be fast enough for dynamic scenes or moving objects, but require special hardware that adds cost and can degrade image quality.  Contrast maximization is image-based and requires no hardware beyond the basic camera itself, but has conventionally required many time steps to achieve satisfactory results. Here we show that deep learning control can achieve the scene-based advantages of contrast optimization at the real-time single frame rates of phase detection. Deep learning further enables dynamic control to image 3D and dynamic scenes in full focus. 

The two key elements of traditional contrast maximization methods are the evaluation metric and the search strategy~\cite{yao2006evaluation}. Evaluation metrics are hand-crafted features that measure the focal degree of an image, and they determine the quality of the captured images. Once the evaluation metric is given, the task of a search strategy is to maximize the metric by efficiently locating the optimal focus position. Considerable investigation has been conducted on these two elements in the literature~\cite{yao2006evaluation, kehtarnavaz2003development, he2003modified, guo2018fast, gamadia2012filter, yousefi2011new, gamadia2010performance}. However, obvious disadvantages exist in these two elements: 1) the evaluation metric itself provides no direct information about the focal state, and thus focal state evaluation requires a large number of comparisons and time steps; and 2) the travel distance in each time step is pre-determined by the search method. In contrast, human can instantly tell from a single image whether a minor or considerable adjustment in focus is required. This gap naturally motivates the deployment of machine learning (ML) techniques in focus control~\cite{park2008fast,chen2010passive, han2011novel, mir2015autofocus, jiang2018transform, wei2018neural, guojin2010image}.

Instead of using explicit evaluation metrics, ML-based methods explore the intrinsic correlation between the image and the focus position. However, most of these methods still act on hand-crafted features extracted from the image. The direct mapping between the image and the focus position remains under-exploited. Here, we propose a deep learning solution to the AF problem and demonstrate that learned focus metrics and optimization methods can substantially improve focus speed and image quality relative to traditional methods.

Traditional AF assumes that there exists a globally optimal focus position. The real world, however, is three-dimensional, suggesting that there is no such global optimum. This issue is often addressed by using expert systems to select a region of interest for focus evaluation~\cite{han2011novel, lee2008enhanced,rahman2008real} or by requesting user input to determine the region of interest. In fact, as the capability of the modern cameras has been increased by the development of hardware, such as fast control module (e.g., voice-coil motor) and fast processing units (e.g., neural computing chips), instead of just recording the instantaneous data, cameras can further estimate the scene from the data and produce an image that is more informative. One of such productions is the all-in-focus image which addresses the absence of optimal focus by fusing multiple frames.

Conventional all-in-focus images are generated from focus stacking, which assumes a static scene and requires capturing a sequence of frames with different focus positions. Even within this setting, the strategy to obtain the focus stack can be improved so that only informative frames are captured. This problem is further complicated for dynamic scenes. In this scenario, the concept of AF is extended to the generation of a focus trajectory to capture the scene so that an image processing can produce all-in-focus video. Here we demonstrate that the proposed deep learning AF methods can be applied to efficiently capture frames for all-in-focus image generation.

The main contributions of this paper are:
\begin{itemize}
\item We demonstrate that deep learning can be used to evaluate distance from the best focus position from a single defocused image. 
Using this network, we build an AF pipeline that outperforms the existing methods in both focus speed and image quality without the sacrifice in computation burden.

\item We develop a new strategy to produce all-in-focus images by integrating the focus control module. This strategy optimizes the image quality in both static and dynamic scenes.

\end{itemize}

The rest of the paper is organized as follows: Section \ref{sec:related work} reviews the previous related work. Section \ref{sec:defous_model} describes the camera defocus model. Section \ref{sec:af-pipeline} and section \ref{sec:all-in-focus} describe the proposed AF pipeline and the focus strategy in all-in-focus imaging. We show implementation results in Section \ref{sec:experiment} and conclude the paper in Section \ref{sec:conclusion}. 

\section{Related Work}
\label{sec:related work}

\subsection{Traditional AF}
Traditional AF systems evaluate the focal state using an evaluation metric. The goal of AF is to maximize this metric. Many different metrics have been considered. The most commonly used metrics are gradient-based, these include absolute gradient, squared gradient, Laplacian filter, Tenengrad function, Brenner function, etc.~\cite{santos1997evaluation, yao2006evaluation}. Other
evaluation metrics can be divided into 4 categories: correlation-based, statistic-based, transform-based and edge-based metrics~\cite{yao2006evaluation}. Correlation-based metrics evaluate the correlation among adjacent pixels, such as autocorrelation, standard deviation~\cite{santos1997evaluation} and joint density function~\cite{yousefi2011new}. Statistic-based metrics derive from the statistics of an image, such as entropy~\cite{santos1997evaluation}, gray level variance~\cite{yao2006evaluation} and histogram~\cite{yao2006evaluation, guo2018fast}. Transform-based metrics analyze the frequency components of the image, and the edge-based metrics exploit the edge information~\cite{yao2006evaluation}.

Traditional systems apply a search strategy to maximize the AF metric. Ideally, such strategies should use as few steps as possible. Popular strategies include  Fibonacci search~\cite{krotkov1988focusing},  rule-based search~\cite{kehtarnavaz2003development} and  hill-climbing search~\cite{he2003modified}. Fibonacci search consecutively narrows the search interval until a desired accuracy is achieved. The number of time steps depends only on the focus range, but it typically requires a long total travel distance.  Rule-based search determines the travel distance in each time step by estimating the distance from the optimal focus. Compared to the Fibonacci search,  rule-based search requires shorter travel distance but more time steps.  Hill-climbing consists of two stages: out-of-focus region search and focus region search. Its performance highly depends on the choice of parameters. A detailed review of these strategies and their variations can be found in \cite{yao2006evaluation}. In general, in order to find the optimal focus position, a large number of time steps are required, which is time- and energy-consuming. To address this issue, curve fitting methods were proposed. A curve fitting method assumes an optical defocus fitting model, e.g. ROL~\cite{yazdanfar2008simple} and BPIC~\cite{wu2012bilateral}, which describes the pattern of the metric with respect to the change of the focus position. Typically only 4-7 time steps are required in curve fitting methods~\cite{wang2018fast}.

The disadvantage of the contrast maximization, time delay, results from the indirect nature of the evaluation metric and the low efficiency of the search strategy. No direct information about the focus position can be learned from the evaluation metric itself. Instead only by a large number of comparisons can one evaluate the focal state, and the optimal focus position cannot be confirmed until the whole focus range has been traveled. The travel distance in each time step is pre-determined by the search strategy. Although some methods such as rule-based search and hill-climbing search adopt an adaptive travel scheme, the travel distance still depends on the strategy's built-in parameters. 

\subsection{ML-based AF}
\label{sec:related work ml}
ML-based AF dates back to 1992 when Hovela {\it et al.}~\cite{hovela1992learning} proposed to use a single perceptron to control the focus module. In the past decade, many other ML techniques have been utilized in AF. Chen {\it et al.}~\cite{chen2010passive} proposed a self-organizing model to predict the focus position from three defocused images. Han {\it et al.}~\cite{han2011novel} proposed a one-nearest neighbor solution. This solution used the focus value increment ratio (FVIR) as the feature which was less dependent on object and illuminance variant. Park {\it et al.}~\cite{park2008fast} proposed a neural network solution that predicted the focus position from the two images at both end of the focus range. These three methods converted the AF to a classification problem, so the focal resolution of the system depended on the number of the pre-defined focus positions. They also required capturing several images at fixed focus positions. Mir {\it et al.}~\cite{mir2015autofocus} proposed an advanced version of the rule-based method using a decision tree, and, compared to traditional methods, more features were considered in the AF process. Although this method was more efficient than the rule-based method, however, it still required more time steps than the Fibonacci search and the hill-climbing search. Chen {\it et al.}~\cite{guojin2010image} proposed to evaluate the quality of the image with a neural network by categorizing the image into three quality levels. All the methods reviewed above operate on the features level, which requires hand-crafted feature design and data processing. In contrast, human mind does not need to “process” the image to evaluate its focal state. An intelligent AF system should similarly achieve optimal focus in only a few, or even one, time step by analyzing the defocused image.

In recent years, deep learning techniques, especially convolutional neural networks (CNN), have been applied to many computer vision tasks, such as image classification~\cite{krizhevsky2012imagenet}, image quality assessment~\cite{Kang_2014_CVPR}, scene recognition~\cite{zhou2014learning}, etc. However, AF, although closely related to image analysis, has not benefited from such techniques. In this paper, a CNN-based AF pipeline is proposed which predicts the optimal focus position directly from the defocused image. Similar ideas have been applied to biomedical imaging~\cite{wei2018neural, jiang2018transform} and digital holography~\cite{ren2018learning}, but, to the best of our knowledge, no CNN-based AF method has been proposed for digital cameras.


\subsection{Focus Stacking}

Cameras have limited Depth of Focus (DoF), thus individual frames cannot be clear everywhere. Focus stacking is one of the deblurring techniques that compensate for the limited DoF by fusing a stack of frames. Frames in the focus stack are captured with different focus settings, thus complimentary clean perception of the scene can be found across the stack. Frame fusion in focus stacking has been well studied in both academia and industry, and its performance has been enhanced by deep learning techniques~\cite{liu2017multi,du2017image, tang2018pixel, guo2018fully}, whereas the strategy to obtain the focus stack received less attention. 

Traditionally a stack of focus is sampled by sweeping over the depth range in uniform steps. Hasinoff {\it et al.}~\cite{hasinoff2009time} discussed the minimum number of frames given the configuration of the camera. 
Considering that the number of frames should also depend on the scene, scene-adaptive strategies were proposed. Vaquero {\it et al.}~\cite{vaquero2011generalized} proposed a 2-step method to select the focus position. The entire focus range was first scanned in a low-resolution fashion. Block-wisely statistical analysis was applied to each captured frame to estimate the focal state of each block. A minimum number of high-resolution images were then captured, ensuring that each block was in focus in at least one frame. This method was further improved by Choi {\it et al.}~\cite{choi2017improved}. The focal state estimation was enhanced by standard depth of field equations, and an explicit block-wise depth estimation was used to improve the selection of the focus position. In these scene-adaptive methods, scanning the entire focus range is required before the focus position for the stack of frames can be determined. Li {\it et al.} \cite{li2018scene} proposed a round-trip acquisition mechanism. The focus position was determined online by constantly estimating the distribution of the focus positions of the entire scene. Although this method aimed to decrease the number of captured frames, the entire focus range still had to be scanned twice, and each capturing required estimating two depth maps and a current all-in-focus image.

The basic assumption in focus stacking is that the scene is static, so the aforementioned methods concentrate on minimizing the number of frames that fully cover the scene. However, the real world is not static.  In order to generate an image or video with as much information as possible, a focus strategy must analyze the previous frames and determine the next focus setting that maximize the quality of the image. In Section \ref{sec:all-in-focus}, we discuss how to develop such a strategy.

\section{Image defocus model}
\label{sec:defous_model}
\begin{figure}[t]
  \centering
  \subfigure[In-focus image]{
    \label{fig:example:infocus} 
    \includegraphics[width=1.6in]{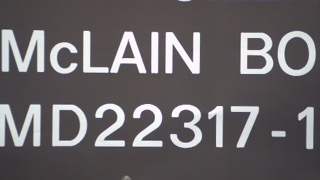}}
  \hspace{0.1in}
  \subfigure[Defocused image]{
    \label{fig:example:defocus} 
    \includegraphics[width=1.6in]{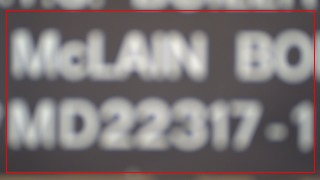}}
  \caption{An example of two images captured by a digital camera. The red bounding box in (b) shows the region captured in the in-focus image, which indicates the magnification change at different focus position.}
  \label{fig:example} 
\end{figure}

Figure \ref{fig:example} shows an example of two images captured by a digital camera. Figure \ref{fig:example:infocus} is the in-focus image, and Figure \ref{fig:example:defocus} is the defocused image. The defocused image can be modelled as generated from the corresponding in-focus image with a blur operation followed by image scaling:
\begin{equation}
Gamma^{-1} (I_d)= imscale(Gamma^{-1}(I_i)*h,\alpha),
\label{eq.model}
\end{equation}
where $I_i$ represents the in-focus image, $I_d$ represents the defocused image, $h$ is the defocus blur filter, $*$ represents convolution operation, $imscale()$ is the image scaling operation, $\alpha$ is the scaling factor and $Gamma^{-1}(x)$ is the inverse process of gamma correction, defined as
\begin{equation}
Gamma^{-1} (I)= I^\gamma.
\end{equation}
This inverse process is to compensate for the non-linear operation, i.e., gamma correction, in the image signal processor (ISP), and $\gamma$ is predefined for a given camera. Note that here we consider gray-scale images.

Traditionally there are two commonly used blur filters: the Gaussian filter and the disk filter. The Gaussian filter is defined as

\begin{equation}
h(x,y)=\frac{1}{2\pi\sigma^2}\exp(-\frac{x^2+y^2}{2\sigma^2}),
\end{equation}
which is determined by the standard deviation $\sigma$, and the disk filter is defined as
\begin{equation}
h(x,y)=\left\{
\begin{aligned}
&\frac{1}{\pi r^2} &&{x^2+y^2\leq r^2}\\
&0  &&{others}\\
\end{aligned}
\right.
,
\end{equation}
which is determined by the defocus radius $r$.

The defocus model depends on the architecture of the camera. Here we present a numerical method to calibrate the model for a given camera. 

\begin{figure}
\centering
\includegraphics[width=0.47\textwidth]{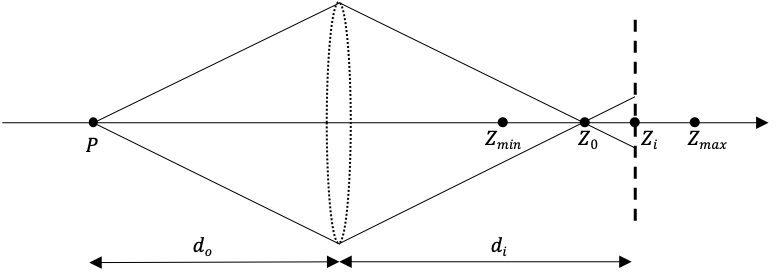}
\caption{\label{fig:optical}Illustration of a lens imaging system.}
\end{figure}

\begin{figure}[htbp]
\centering
\includegraphics[width=0.48\textwidth]{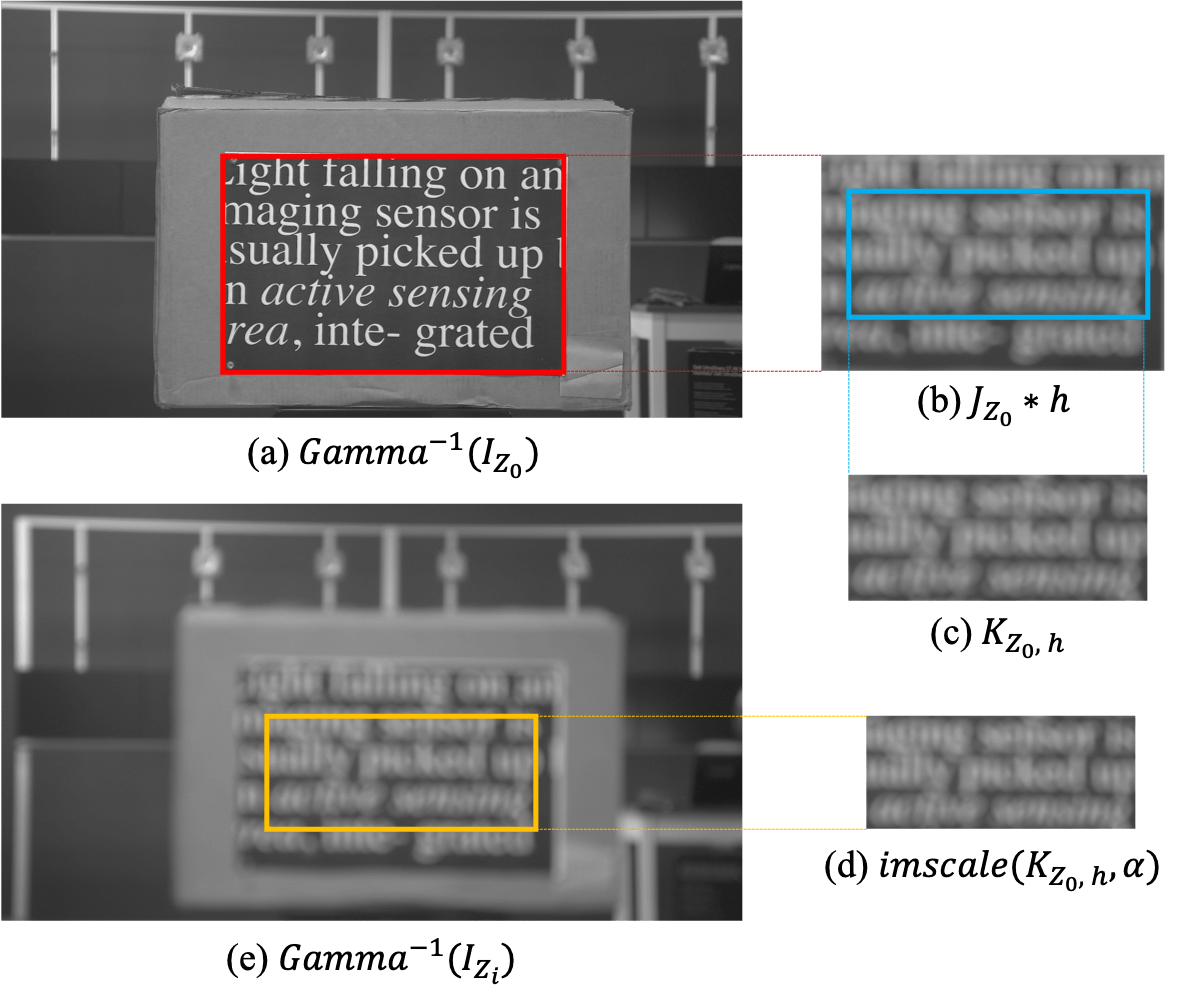}
\caption{\label{fig:process}Illustration of the calibration process.}
\end{figure}

A simplified imaging system is illustrated in Figure \ref{fig:optical}. The light from a point source $P$ converges at $Z_0$ on the optical axis, and the camera sensor array can move in the interval $(Z_{min},Z_{max})$. Then the specification of the defocus model, i.e., the scaling factor $\alpha$ and the blur filter $h$, should be the function of both the optimal focus position $Z_0$ and the position of the sensor array $Z_i$. Let $I_{Z_0}$ and $I_{Z_i}$ denote the images captured at $Z_0$ and $Z_i$. A two-dimensional brute-force search is applied to calibrate the model for each pair of $Z_0$ and $Z_i$. The process is illustrated in Figure \ref{fig:process}:

1. A patch is chosen from $Gamma^{-1}(I_{Z_0})$, denoted as $J_{Z_0}$. All the objects (pixels) in this patch should be in focus. See the red bounding box in Figure \ref{fig:process}(a).

2. For each standard deviation $\sigma$ for Gaussian filter (or defocus radius $r$ for disk filter), compute $J_{Z_0}*h$. See Figure \ref{fig:process}(b).

3. An inner patch is chosen from $J_{Z_0}*h$, denoted as $K_{Z_0,h}$. See the blue bounding box in Figure \ref{fig:process}(b) and Figure \ref{fig:process}(c). This step is to remove the pixels that should have been blurred by the pixels outside the $J_{Z_0}$. 

4. For each scaling factor $\alpha$, generate the scaled image $imscale(K_{Z_0,h},\alpha)$. See Figure \ref{fig:process}(d).

5. A score for each pair of $h$ and $\alpha$ is obtained by computing the maximum normalized cross-correlation between $imscale(K_{Z_0,h},\alpha)$ and all patches that have the same dimension in $Gamma^{-1}(I_{Z_i})$.

6. The $h$ and $\alpha$ corresponding to the maximum score are the calibrated model parameters.

An example of a calibrated model is shown in Figure \ref{fig:calibration}, and how the model is used to generate data is discussed in Section \ref{sec:train_detail}.

\section{Deep Learning AF Pipeline}
\label{sec:af-pipeline}

\begin{figure*}
	\centering
	\subfigure[System Structure\label{fig:af_system_structure}]{\includegraphics[width=0.72\linewidth]{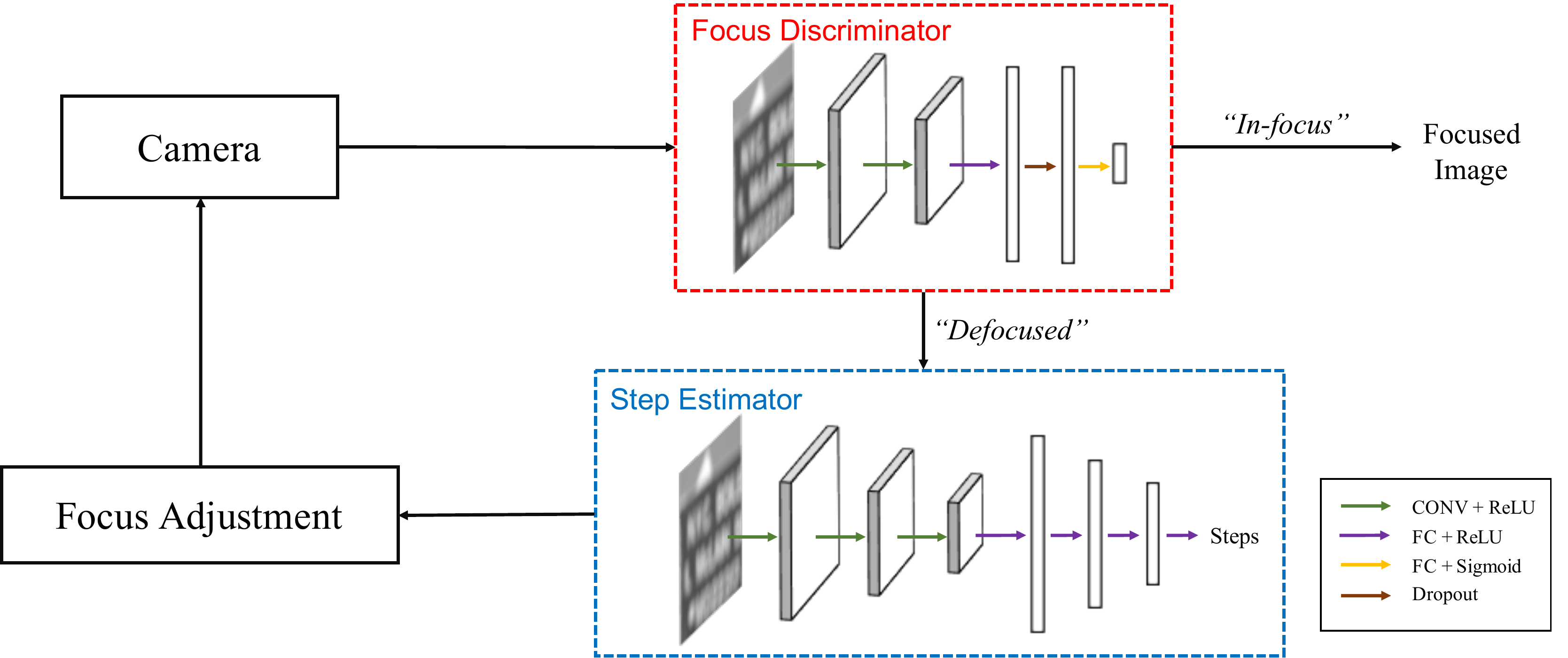}}
	\hspace{0.1in}
	\subfigure[System Implementation\label{fig:af_system_implementation}]{\includegraphics[width=0.23\linewidth]{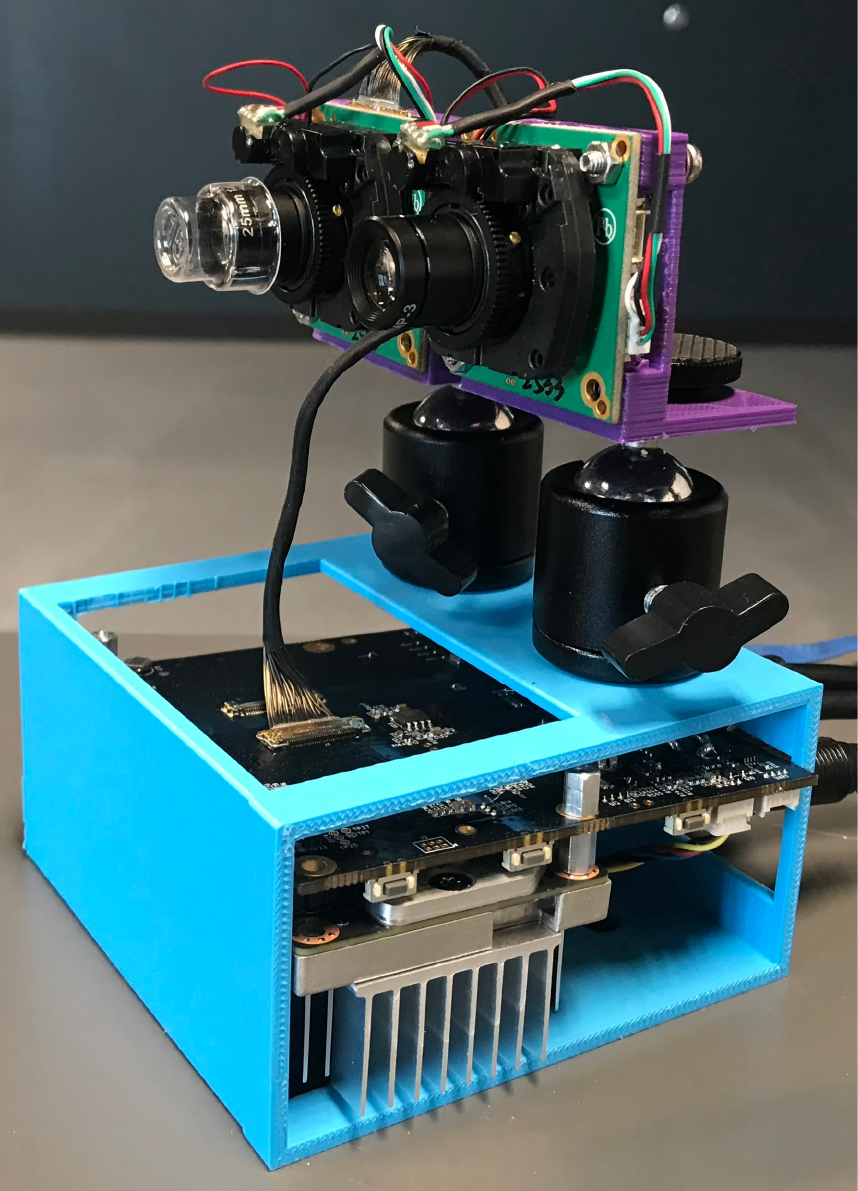}}
	\caption{Overview of the proposed AF pipeline. (a) Input to the discriminator or the estimator is a patch of size $512\times512$ from the image. \textbf{Focus discriminator}: The filter size/number of filters/stride for the two convolutional layers are $8\times8/1/8$ and $8\times8/1/8$. The dimension of the fully-connected layers are 10 and 1. The dropout rate for the dropout layer is 0.5. \textbf{Step estimator}: The filter size/number of filters/stride for the three convolutional layers are $8\times8/4/8$, $4\times4/8/4$ and $4\times4/8/4$. The dimension of the fully-connected layers are 1024, 512, 10 and 1. (b) The network configuration is tested on a camera module with Evetar lens (25mm, F/2.4) and CMOS image sensor (Sony IMX274 4K).}
	\label{fig:AF_example}
\end{figure*}

We restate the disadvantages of the traditional contrast maximization methods here:
\begin{itemize}
\item The evaluation metric itself provides no direct information about the focal state; and
\item The travel distance in each time step is pre-defined by the search strategy.
\end{itemize}
These disadvantages indicate that an intelligent AF system should be able to determine whether a given image is in focus, and, if not, how much travel distance is required to approach the optimal focus. Motivated by these two needs, we propose a deep learning AF pipeline. There are mainly two components in this pipeline: a step estimator and a focus discriminator. The pipeline is illustrated in Figure.~\ref{fig:af_system_structure}. The network configuration in this section has been tested on a camera module with Evetar lens (25mm, F/2.4) and CMOS image sensor (Sony IMX274 4K), shown in Figure.~\ref{fig:af_system_implementation}. The same pipeline applies to different cameras with minor changes in network configuration.

\subsection{Step Estimator}
In traditional AF, the search strategy determines the travel distance in each time step. However, the focal degree of an image reveals the distance from the optimal focus position and should be considered in the strategy. Here we argue that a defocused image should contain sufficient information to estimate the distance (or the number of motor steps) from the optimal focus position, and we build a CNN, named {\it step estimator}, to achieve this purpose.

The structure of the proposed step estimator, denoted as $f_{e}$, is illustrated in the blue bounding box in Figure \ref{fig:af_system_structure}. The input to the network is an image patch of size $512\times512$. The output of the network is the estimated steps from the in-focus position which is a positive scalar without indicating the movement direction:

\begin{equation}
f_e(I_{Z_i})=abs(Z_i-Z_0).
\end{equation}

\subsection{Focus Discriminator}
In traditional AF, the algorithm terminates until the entire focus range has been traveled~\cite{kehtarnavaz2003development} or the minimum number of steps is achieved~\cite{krotkov1988focusing}. This searching manner is inefficient, because it cannot identify the in-focus image during the search process, and it typically requires a system reset at first, thus a mechanism to identify the in-focus image and terminate the algorithm is necessary in an efficient AF system. Here we propose a focus discriminator, denoted as $f_d$, that directly determines if an image is in focus and ends the algorithm when the optimal focus is achieved.

The structure of the discriminator is illustrated in the red bounding box in Figure \ref{fig:af_system_structure}. The input to the network is an image patch of size $512\times512$, and the output of the network is the predicted label of the the image: in-focus or defocus.

\subsection{The Proposed Pipeline}
Algorithm \ref{af algotirhm} shows the proposed AF pipeline in algorithmic form. The algorithm starts with an arbitrary focus position. In each time step, the focus discriminator evaluates the image and stops the algorithm when an in-focus image is captured. Otherwise, the step estimator predicts the travel distance, and up to two candidate (both directions) positions are obtained. The focus position is updated with the one that has better image quality. The comparison between multiple candidate positions is completed by comparing their estimated distance from the optimal focus position. The advantage is that the distance can be utilized in the next time step, which improves the algorithm efficiency.

\begin{algorithm}
\caption{Deep-learning based AF} 
\label{af algotirhm}
\begin{algorithmic}[1]
\REQUIRE ~~\\ 
Camera focus range $(Z_{min},Z_{max})$;\\
Initialized focus position $Z$;\\
\ENSURE ~~\\ 
Optimal focus position;
\STATE Verified $\Leftarrow False$
\WHILE{true}
\IF{Verified $=False$}
\IF{$f_d(I_Z)=1$}
\RETURN Z
\ENDIF
\STATE $Z_d \Leftarrow f_e(I_Z)$
\ENDIF
\IF{$Z+Z_d>Z_{max}$ \AND $Z-Z_d<Z_{min}$}
\STATE $Z \Leftarrow uniform(Z_{min},Z_{max})$, Verified $\Leftarrow False$
\ELSIF{$Z+Z_d<Z_{max}$ \AND $Z-Z_d<Z_{min}$}
\STATE $Z \Leftarrow Z+Z_d$, Verified $\Leftarrow False$
\ELSIF{$Z+Z_d>Z_{max}$ \AND $Z-Z_d>Z_{min}$}
\STATE $Z \Leftarrow Z-Z_d$, Verified $\Leftarrow False$
\ELSE
\STATE ${Z_1}\Leftarrow Z+Z_d$, ${Z_2}\Leftarrow Z-Z_d$
\IF{$f_e(I_{Z_1})<f_e(I_{Z_2})$}
\IF{$f_d(I_{Z_1})=1$}
\RETURN $Z_1$
\ELSE
\STATE $Z \Leftarrow Z_1$, $Z_d \Leftarrow f_e(I_{Z_1})$
\ENDIF
\ELSE
\IF{$f_d(I_{Z_2})=1$}
\RETURN $Z_2$
\ELSE
\STATE $Z \Leftarrow Z_2$, $Z_d \Leftarrow f_e(I_{Z_2})$
\ENDIF
\ENDIF
\STATE Verified $\Leftarrow True$
\ENDIF
\ENDWHILE
\end{algorithmic}
\end{algorithm}

\section{Focus Strategy in All-in-focus Imaging}
\label{sec:all-in-focus}
Obtaining the focus stack is not a trivial problem in producing all-in-focus images, especially when the scene is dynamic. However, strategies discussed in previous work still have large room to improve. Scene-independent strategies always capture the same number of frames, while scene-adaptive algorithms sacrifice the time because the entire focus range needs to be swept more than once, and all these strategies deal with static scenes. 

In view of the recurrent nature of the capture process, the focus position in each time step should be determined by the states of the camera and the scene in previous time steps, while keeping a balance of information acquisition and time budget. Here we propose a new strategy that determines the focus position in an adaptive way, by utilizing the step estimator, that optimizes the quality of the fused all-in-focus images.



\subsection{Focus Strategy}

\begin{figure*}
\centering
\subfigure[The pipeline of the focus strategy.]{\includegraphics[width=0.98\linewidth]{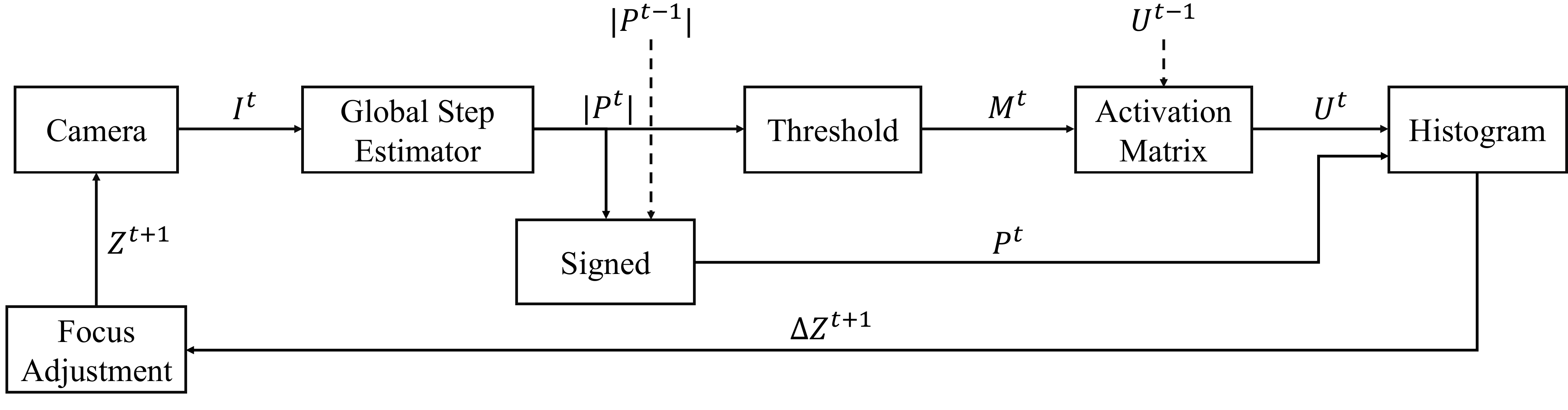}}
\subfigure[Image illustration of the pipeline.]{\includegraphics[width=0.98\linewidth]{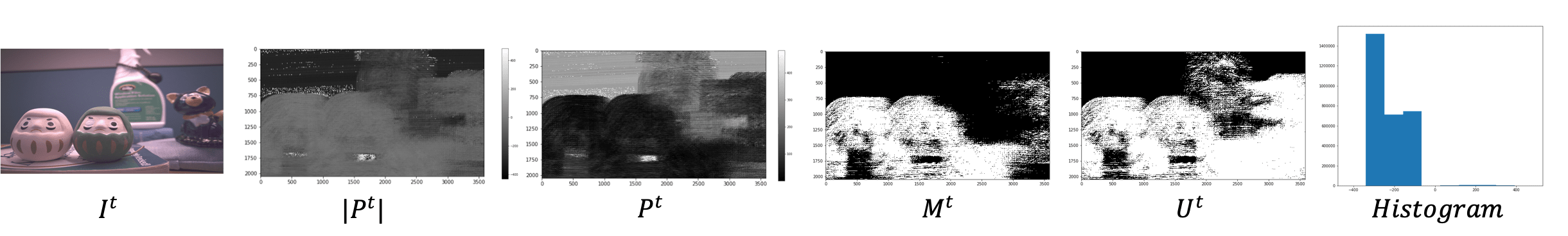}}

\caption{\label{fig:focus_policy} The proposed focus strategy. An image sample is provided to better illustrate the pipeline. The dashed arrow denotes the information from previous time steps.}
\end{figure*}

Our focus strategy is as described in Figure. \ref{fig:focus_policy}. The system starts at an arbitrary focus position $Z^0$ at time $t=0$. 
At time step $t$, the camera captures $I^t$ w.r.t. the last predicted focus position $Z^t$. 

First we compute a focus deviation matrix $P^t$ by applying the step estimator to all the $512\times 512$ patches of $I^t$, and we especially design a \textbf{global step estimator} to accelerate the process (see Section \ref{sec:globalstepest}). At this stage $\lvert P^t \rvert$ composes absolute steps from $Z^t$, where the movement direction is not specified. When $t \geq 1$, we can determine the sign of each element $\lvert p^t \rvert$ in $\lvert P^t \rvert$ with the knowledge of $\lvert p^{t-1} \rvert$, $Z^{t-1}$ and $Z^t$: 

\begin{equation}
p^t = \begin{cases} 
      \lvert p^t \rvert & \lvert \lvert \lvert p^t \rvert - (Z^{t} - Z^{t-1}) \rvert - \lvert p^{t-1} \rvert \rvert \leq \\
      & \lvert \lvert \lvert p^t \rvert + (Z^{t} - Z^{t-1}) \rvert - \lvert p^{t-1} \rvert \rvert\\
      -\lvert p^t \rvert & otherwise
   \end{cases}
   \label{eq:q}
\end{equation}

\noindent hence the signed matrix $P^t$ can be constructed.

Next we verify the focal state of each individual patch. A binary in-focus mask $M^t$ is derived by the equation $M^t = Boolean(\lvert P^t \rvert \dot \leq \sigma_{Z^t})$, where $\dot \leq$ is an element-wise operation, and $\sigma_{Z^t}$ is the threshold determined by the DoF. The subscript implies that the threshold varies depending on $Z^t$.

The mask is then integrated as a component of the \textbf{activation matrix} $U^t$:
\begin{equation}
    U^t = f(M^{0}, M^{1}, \dots, M^t),
    \label{eq: activation matrix}
\end{equation}
which is also a binary matrix. Elements marked as $0$s indicate the pending-processing area, from which we derive the next step, while $1$s indicate where currently no action is required. $f$ is designed to aggregate in-focus mask information temporally, which behaves distinctively under static or dynamic conditions (see Section \ref{sec: ut generation}). In practice, Eq. \ref{eq: activation matrix} may be simplified as:

\begin{equation}
    U^t = f(M^t, U^{t-1}).
\end{equation}

Given $U^t$, we draw a histogram of step values from the area of interest in the final step, namely $P^t \odot \overline{U^t}$, where $\odot$ denotes element-wise multiplication and $\overline{U^t}$ denotes the complement of $U^t$. The center of the bin with most pixels is selected as $\Delta Z^{t+1}$, where next focus position is derived by $Z^{t+1} = Z^t + \Delta Z^{t+1}$.

Static and dynamic all-in-focus imaging largely share the same pipeline. The difference is the termination. In static cases, the capture process is terminated when more captured frames do not improve the quality of the fused image, which typically happens when

\begin{itemize}
    \item the remaining regions are out of the focus range; or
    \item the next predicted focus position is within the DoF of any previous focus positions, which may happen when noise level is high and the system tries to "refine" the noisy regions.
\end{itemize}

\subsection{Global Step Estimator} 
\label{sec:globalstepest}
The global step estimator inherits the structure and parameters from the step estimator, and we achieve this by converting the trained step estimator to a fully convolutional network which takes in any image larger than $512\times512$ and outputs distances from optimal focus for all the $512\times512$ patches in the image. The convolutional layers are replaced by dilated convolutional layers, and the the fully-connected layers are replaced by convolutional layers with filter size 1. The detailed network configuration is summarized in Table \ref{tab:pixel_wise_convert}. By this conversion, the estimator and the global estimator becomes one network. 

\begin{table}[htbp]
  \centering
  \caption{\rm The conversion from the step estimator to the global step estimator. The three parameters in Conv() represent the number of filters, filter size and stride of the convolutional layer. The parameter in FC() represents the output dimension of the fully-connected layer.}
    \begin{tabular}{cc}
    \toprule
    Step estimator & Global step estimator \\
    \midrule
    Conv(4,8,8) & Conv(4,8,1) \\
    \midrule
    Conv(8,4,4) & Conv(8,4,1,dilation=8)  \\
    \midrule
    Conv(8,4,4) & Conv(8,4,1,dilation=32)  \\
    \midrule
    Flatten + FC(1024) & Conv(1024,4,1,dilation=128)  \\
    \midrule
    FC(512) & Conv(512,1,1) \\
    \midrule
    FC(10) & Conv(10,1,1) \\
    \midrule
    FC(1) & Conv(1,1,1) \\
    \bottomrule
    \end{tabular}%
  \label{tab:pixel_wise_convert}%
\end{table}%

\subsection{Activation Matrix}
\label{sec: ut generation}

The goal of constructing $U^t$ is to spot the region that needs processing at next time step. The function $f$ accordingly incorporates previous focal states, and its realization depends on the tasks. 

In static cases, when the goal is to capture frames so that every region is clear in at least one frame, the activation matrix can be updated by accumulating the regions that have already been in-focused:
\begin{equation}
    U^t = U^{t-1} \lor M^t
\end{equation}
where $\lor$ denotes logical or.

In dynamic cases, accumulating all previous frames is not reasonable. In fact, considering that there exists strong similarity between consecutive frames, we can compute the activation matrix by just accumulating the previous $k$ frames. $k$ is a hyper-parameter and should be determined by the scene. When there are large disparity, i.e. the scene is strongly dynamic, $k$ should be small. We can further adopt a greedy strategy by simply defining $k=0$ and $U^t = M^t$. The strategy for dynamic cases is summarized in Algorithm \ref{all-in-focus algotirhm}.


\begin{algorithm}
\caption{Focus Strategy for Dynamic All-in-focus Imaging. $f_g$ is the global step estimator.} 
\label{all-in-focus algotirhm}
\begin{algorithmic}[0]
\STATE Initialize focus position $Z^0$;\\
\STATE Specify the hyper-parameter $k$;\\
\STATE $t = 0$
\FOR{number of frames}
\STATE $\bullet$ Capture a frame $I^t$ at focus position $Z^t$.
\STATE $\bullet$ Compute the absolute deviation matrix $|P^t|$:
\[|P^t|=f_g(I^t).\]
\STATE $\bullet$ Compute the deviation matrix $P^t$. If $t>0$, compute $P^t$ following Eq. \ref{eq:q}; if $t=0$, $P^t = \lvert P^t \rvert$.

\STATE $\bullet$ Compute the in-focus mask $M^t$:
\[M^t =Boolean(\lvert P^t \rvert \dot \leq \sigma_{Z^t}).\]
\STATE $\bullet$ Compute the activation matrix mask $U^t$:
\[U^t = M^t\lor M^{t-1}\lor\dots\lor M^{t-k}.\]
\STATE $\bullet$ Draw a histogram from $P^t \odot \overline{U^t}$ and determine travel distance $\Delta Z^{t+1}$ by finding the bin with most pixels.
\STATE $\bullet$ $t = t+1$, $Z^{t+1} = Z^t+\Delta Z^{t+1}$
\ENDFOR
\end{algorithmic}
\end{algorithm}

\section{System Implementation and Experiments}
\label{sec:experiment}
We implemented both the AF pipeline and an all-in-focus imaging system. Through experiments, we demonstrate that the proposed methods outperforms the existing methods in both image quality and efficiency.

\subsection{AF pipeline}
\subsubsection{Defocus Model Calibration}
\label{sec:model_calibration}
\begin{figure}
	\centering
	\subfigure[System Structure\label{fig:calibration_r}]{\includegraphics[width=0.49\linewidth]{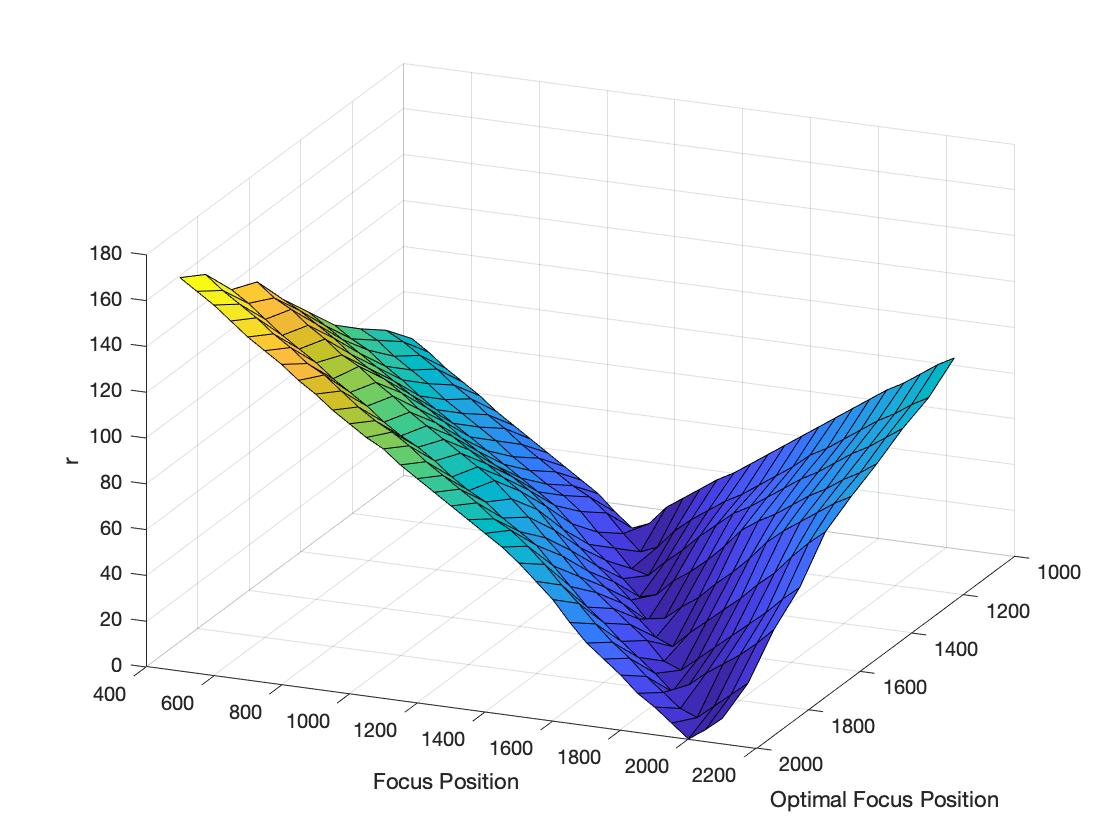}}
	\subfigure[System Implementation\label{fig:calibration_alpha}]{\includegraphics[width=0.49\linewidth]{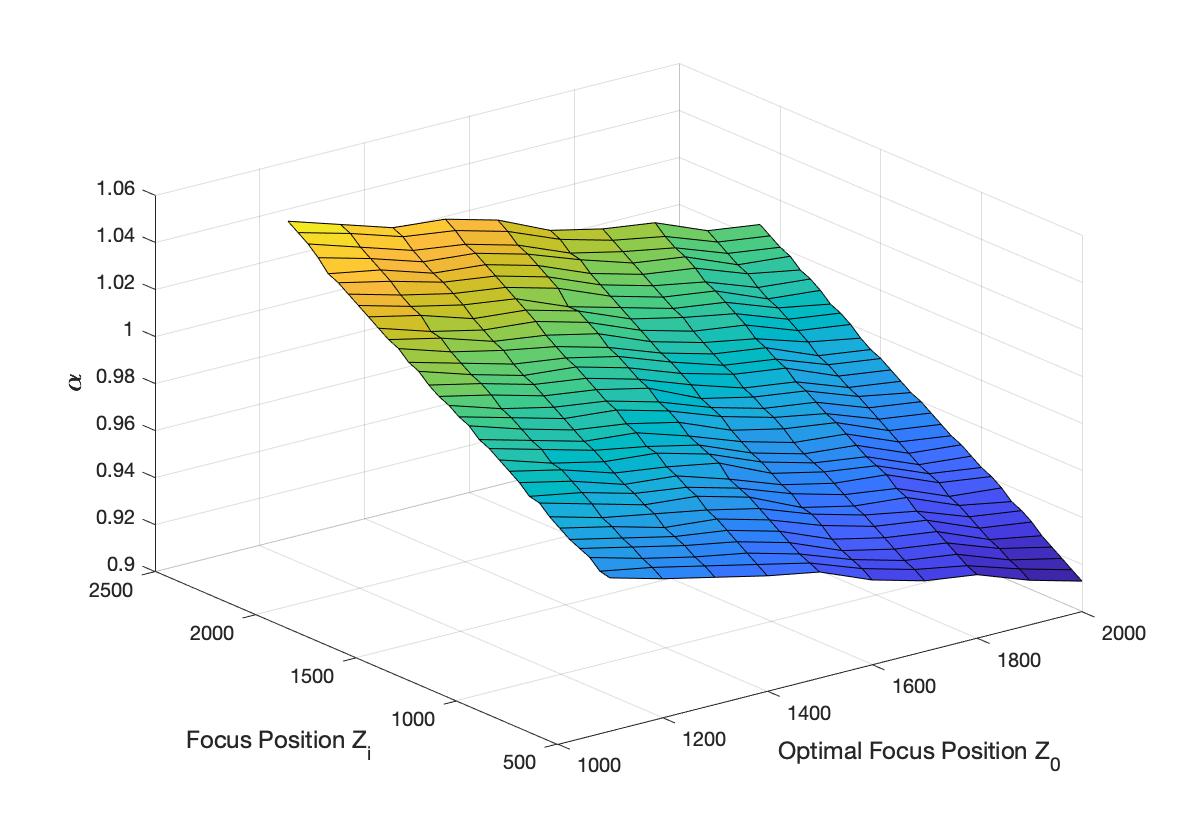}}
	\caption{The calibrated model. (a) shows the defocus radius $r$, and (b) shows the scaling factor $\alpha$. Both $r$ and $\alpha$ are functions of the focus position and the optimal focus position.}
	\label{fig:calibration}
\end{figure}

We used the system in Figure \ref{fig:af_system_implementation} to validate the proposed AF pipeline. The camera was driven by NVIDIA Jetson TX1. There were 2200 stepping motor positions, and we calibrated the system in the interval $(1050,2100)$ which covered indoor scenarios (approximately 40 centimeters to 10 meters). The disk blur filter was used, and the calibration was performed every 50 steps due to the time limit. The objects were illuminated by fluorescent lamp light during the calibration. Figure \ref{fig:calibration} shows the calibrated model. Note that the maximum defocus radius is larger than 160 pixels. Thus we considered a patch size larger than 320 for better AF performance, and $512\times512$ was chosen based on both computation efficiency and performance.

\subsubsection{Training Details}
\label{sec:train_detail}
To train the estimator and the discriminator, we simulated defocused images from existing image dataset, and we regarded images in those dataset as in-focus images. For each image, we randomly selected its in-focus position, $Z_0$, and defocus position, $Z_i$. Then a defocused image was generated from Eq. \ref{eq.model}, where $h$ and $\alpha$ had been calibrated. For the estimator, the training label was $abs(Z_0-Z)$; and for the discriminator, the label was 1 if $Z_0 = Z_i$, and 0 otherwise. We generated 20000 training images in total from DIV2K dataset \cite{Agustsson_2017_CVPR_Workshops} and CLIC dataset~\cite{clic}.

All the networks were trained using Tensorflow~\cite{tensorflow2015-whitepaper} on Nvidia GeForce GTX 1070. The loss function was the average mean square error over all the training data, and the networks were trained with Adam optimizer~\cite{kingma2014adam}. For the estimator, we used a batch size of 32 and initial learning rate 0.001. The network was trained for two hours over 300 epochs. For the discriminator, we used a batch size of 128 and initial learning rate 0.001. The network was trained for one hours over 100 epochs.

\subsubsection{Experimental Results}
\begin{figure}
\centering
\includegraphics[width=0.46\textwidth]{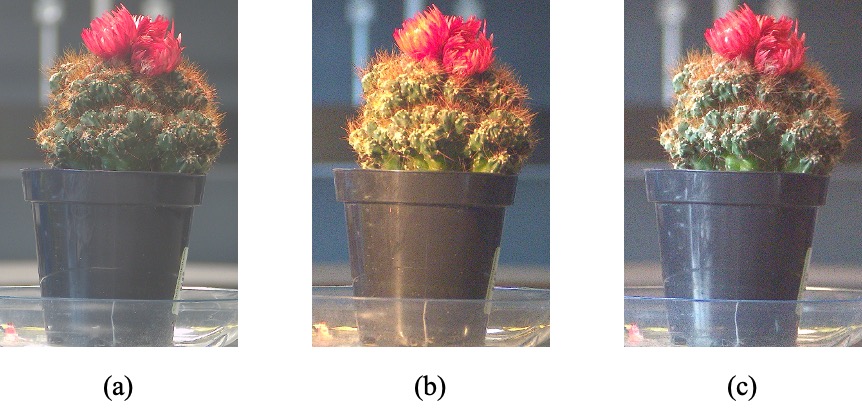}
\caption{\label{fig:illumination} Three different light sources: (a) fluorescent lamp, (b) incandescent bulb light, and (c) natural light.} 
\end{figure}

\textbf{Effectiveness and Robustness}
First we validate the effectiveness and robustness of the system. The proposed AF method was used to focus on objects with different light sources (Figure \ref{fig:illumination}). The objects were installed at different distances, and the algorithm was initialized at four different starting positions: 1100, 1400, 1700 and 2000. The results are shown in Table \ref{tab:illumination}. Note that the DoF for this camera is around 20 steps.

Although the defocus model was calibrated under fluorescent lamp, the proposed method achieved optimal focus for all light sources. This demonstrates that the network can be effectively applied to different scenarios once it has been trained. Besides, the performance of the proposed method was not affected by the staring position.

It is worth noting that, as shown in Figure \ref{fig:illumination}, the noise level under incandescent bulb light and natural light in the experiments were much higher than under fluorescent lamp, which demonstrates the method's robustness to noise.

\begin{table}[htbp]
  \centering
  \caption{\rm Results of the proposed AF method with different light sources. The number in the parentheses is the groundtruth optimal focus position of the object. Although the model is calibrated every 50 steps, the AF performs in continuous manner.}
    \begin{tabular}{cccccc}
    \toprule
    Object & Light & 1100  & 1400  & 1700  & 2000 \\
    \midrule
          & Fluorescent Lamp & 1611  & 1607  & 1599  & 1619 \\
    Plant (1600) & Bulb Light & 1610  & 1582  & 1598  & 1609 \\
      & Natural Light & 1613  & 1589  & 1587  & 1605 \\
    \midrule
          & Fluorescent Lamp & 1813  & 1813  & 1803  & 1808 \\
    Toy (1800)   & Bulb Light & 1820  & 1832  & 1823  & 1830 \\
        & Natural Light & 1801  & 1832  & 1812  & 1816 \\
    \midrule
          & Fluorescent Lamp & 1298  & 1306  & 1301  & 1321 \\
    Books (1300) & Bulb Light & 1358  & 1304  & 1289  & 1318 \\
          & Natural Light & 1305  & 1292  & 1325  & 1353 \\
    \midrule
          & Fluorescent Lamp & 1453  & 1464  & 1430  & 1454 \\
    Doll (1450)  & Bulb Light & 1461  & 1465  & 1472  & 1471 \\
       & Natural Light & 1448  & 1466  & 1481  & 1452 \\
    \midrule
          & Fluorescent Lamp & 1740  & 1746  & 1731  & 1756 \\
    Bottle (1750) & Bulb Light & 1741  & 1735  & 1738  & 1763 \\
       & Natural Light & 1749  & 1756  & 1732  & 1741 \\
    \bottomrule
    \end{tabular}%
  \label{tab:illumination}%
\end{table}%

\textbf{Efficiency Comparison with Existing Methods} As discussed in Section \ref{sec:related work ml}, the existing ML-based AF either converts the AF to a classification problem~\cite{park2008fast,chen2010passive,han2011novel}, which limits the image quality, or still adopts the traditional search manner and takes longer time than the Fibonacci search~\cite{mir2015autofocus}, so we limit our comparisons to traditional AF methods.

\begin{figure}
	\centering
	\subfigure{\includegraphics[width=0.24\linewidth]{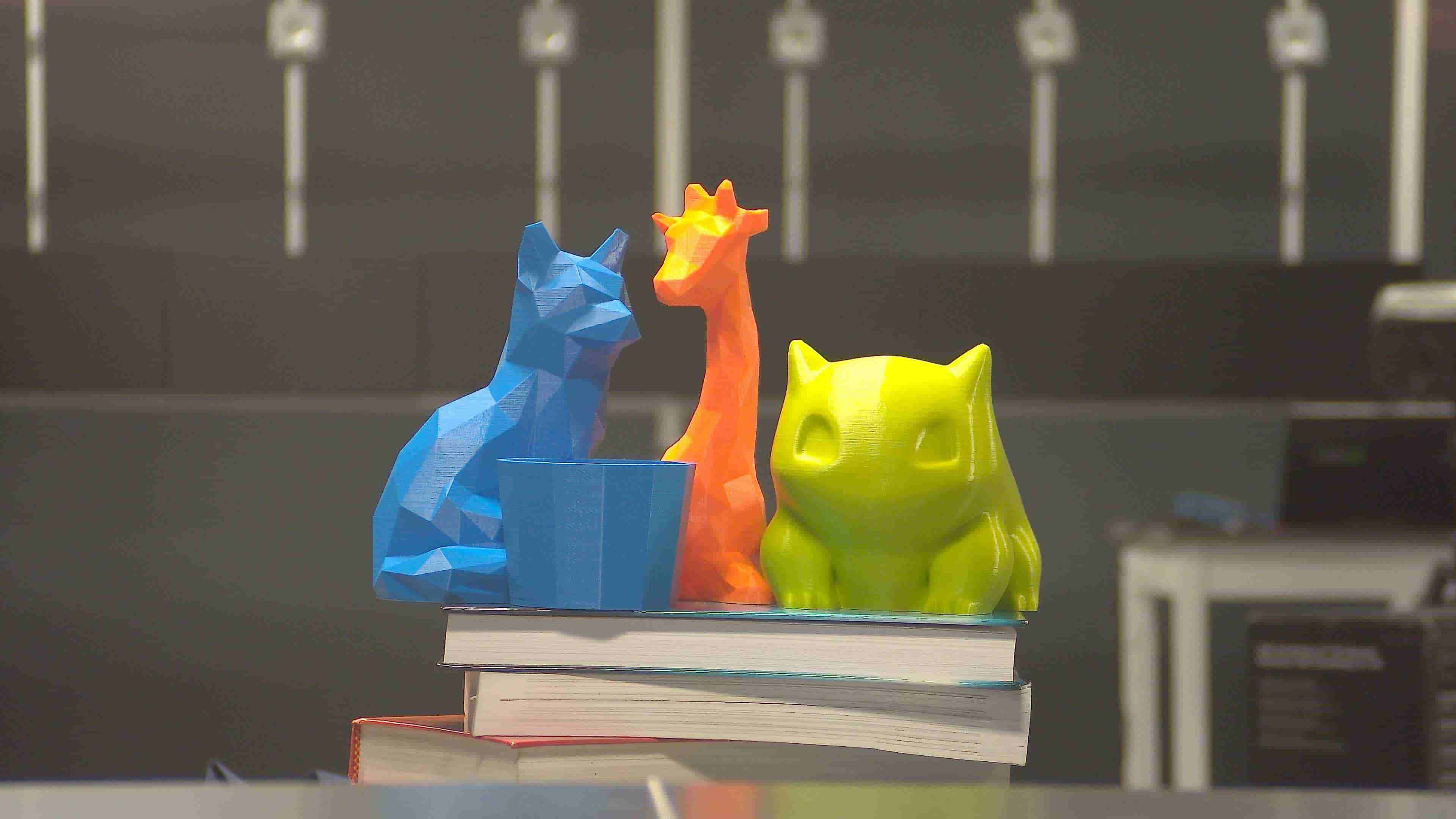}}
	\subfigure{\includegraphics[width=0.24\linewidth]{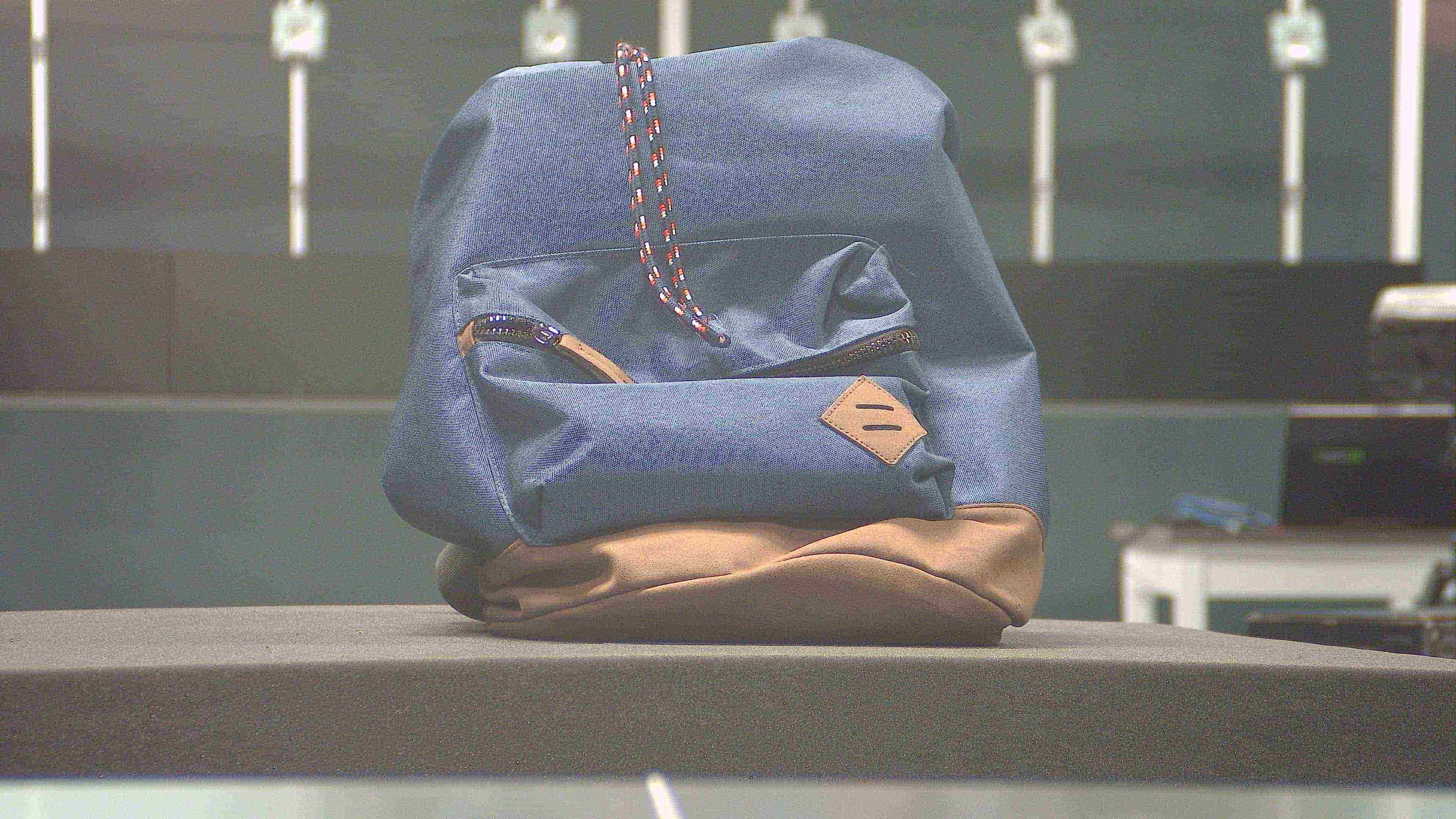}}
	\subfigure{\includegraphics[width=0.24\linewidth]{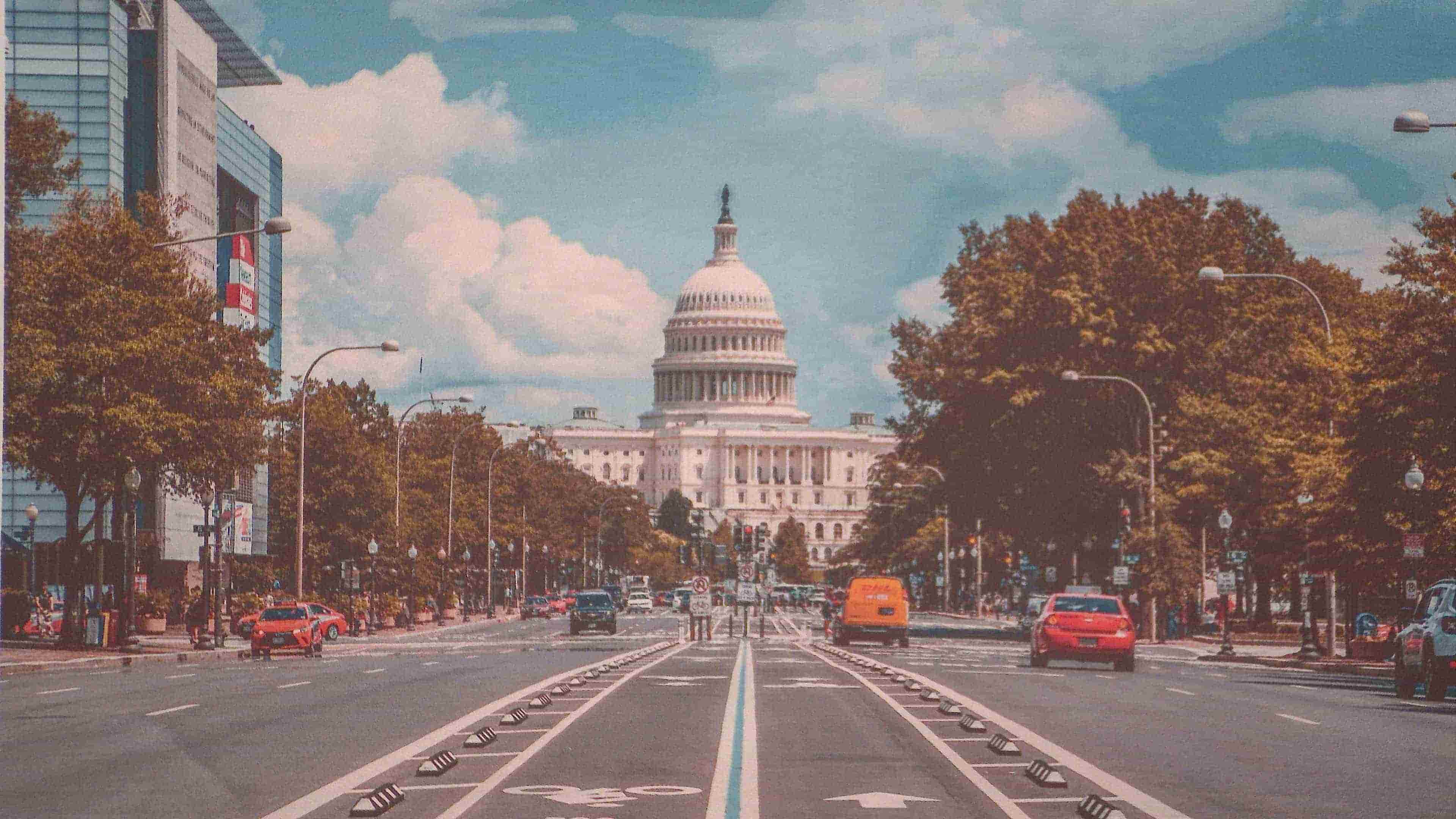}}
	\subfigure{\includegraphics[width=0.24\linewidth]{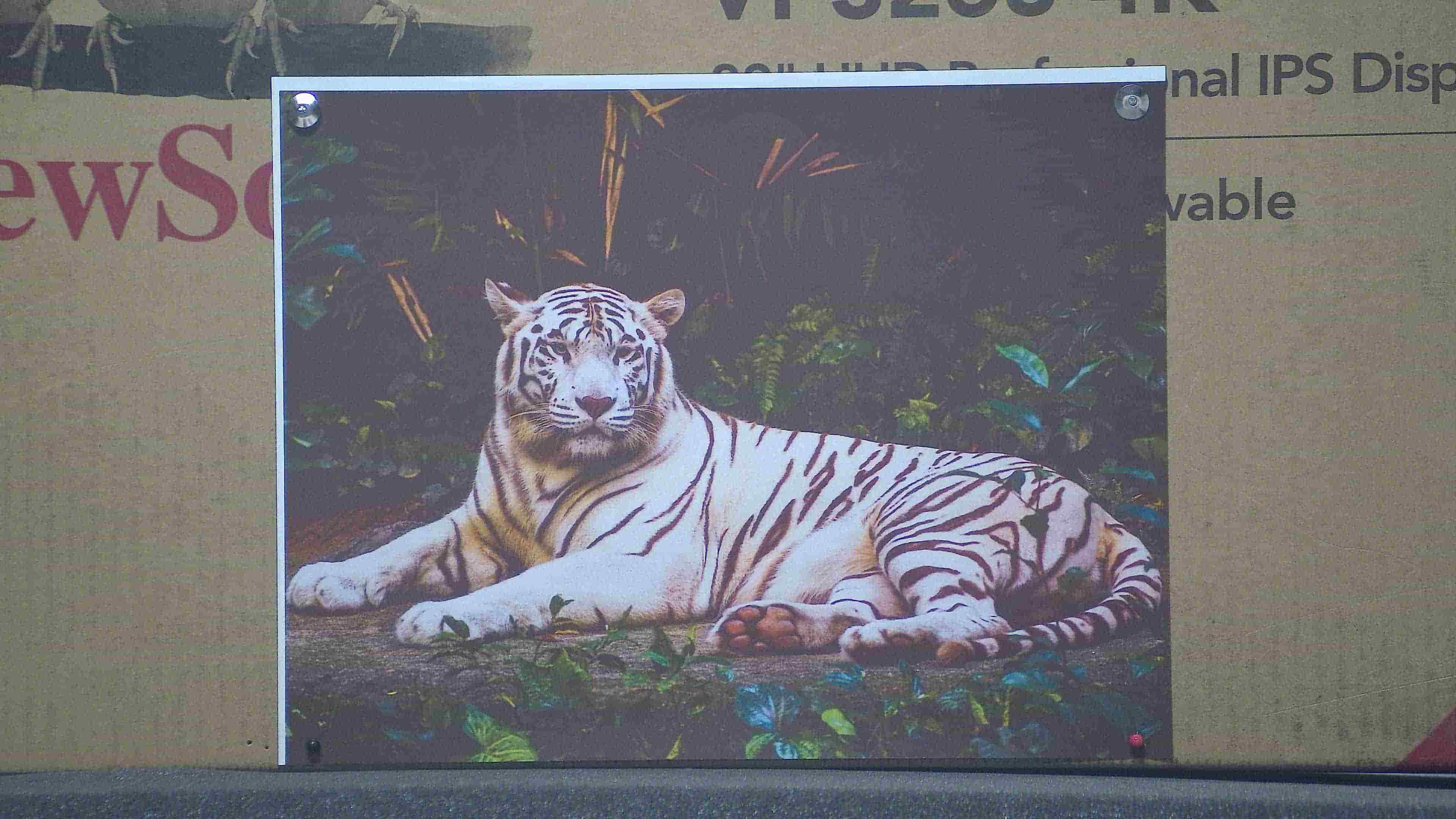}}
	\caption{Object and image samples used in the experiments.}
	\label{fig:af_object_samples}
\end{figure}

We implemented Fibonacci search~\cite{krotkov1988focusing} and rule-based search~\cite{kehtarnavaz2003development} on our system, and we selected the Tenengrad \cite{yao2006evaluation} as the evaluation metric for these two strategies. In the rule-based search, \emph{Initial}, \emph{Coarse}, \emph{Fine} and \emph{Mid} were set to be 10, 40, 10 and 30. The three methods were used to focus on 10 different images and 8 different objects. Some of the targets are shown in Figure~\ref{fig:af_object_samples}. Every image or object was installed at a random distance from the camera, and we recorded the number of time steps for each method. The results are shown in Table \ref{tab:image} and Table \ref{tab:table_object}. 

Different from traditional AF which determines the optimal focus position by a series of comparisons, the proposed pipeline directly makes decisions on the images by introducing a focus discriminator. This image-based decision allows the algorithm to return the optimal focus position as soon as it is obtained. Therefore, compared to the traditional search methods which typically require over 10 time steps, the proposed method can achieve optimal within only a few, or even one, time steps. Another advantage is that this method does not require a specific starting position. This, compared to the existing methods, both learning based and non-learning based, is more efficient and can be better applied in different applications, such as dynamic focus control.

In Table \ref{tab:time_compare}, we also compare the actual time consumption between the proposed method and the traditional methods by reporting the computation time for each iteration. Although deep learning is generally associated with heavy computation, the proposed method did not suffer from it. In fact, even without GPU acceleration, the proposed method was still comparable to traditional methods. That being said, our method benefits from deep learning's strength in data analysis without the sacrifice in computation efficiency.


\begin{table}[htbp]
  \centering
  \caption{\rm Average time consumption by different components in AF systems based on $512\times512$ image patches. The "Proposed" is the total time of the step estimator and the focus discriminator. The "GPU" stands for GeForce GTX 1070, and the "CPU" stands for 2.7 GHz Intel Core i5.}
    \begin{tabular}{cccccc}
    \toprule
    Hardware & Traditional &  Estimator & Discriminator & Proposed \\
    \midrule
    GPU & 0.00364s & 0.00120s  & 0.00094s & \textbf{0.00214s}   \\
    \midrule
    CPU  & 0.00389s & 0.00211s  & 0.00145s  & \textbf{0.00356s}  \\
    \bottomrule
    \end{tabular}%
  \label{tab:time_compare}%
\end{table}%

\begin{table*}
  \centering
  \caption{ {\rm Number of time steps for different images. The number in the parentheses is the groundtruth focus position.} }
    \begin{tabular}{ccccccc}
    \toprule
    Image & Rule-based & Fibonacci & Proposed-1100 & Proposed-1400 & Proposed-1700 & Proposed-2000 \\
    \midrule
    Street1 (1570) & 63    & 13    & 1     & 1     & 2     & 1 \\
    \midrule
    Street2 (1310) & 44    & 13    & 1     & 2     & 3     & 3 \\
    \midrule
    Building1 (1150) & 37    & 13    & 1     & 2     & 2     & 3 \\
    \midrule
    Building2 (1850) & 71    & 13    & 2     & 1     & 1     & 1 \\
    \midrule
    Building3 (1590) & 65    & 13    & 1     & 1     & 2     & 3 \\
    \midrule
    Tiger (1390) & 50    & 13    & 1     & 0     & 2     & 2 \\
    \midrule
    Eagle (1280) & 46    & 13    & 1     & 2     & 1     & 2 \\
    \midrule
    Scenery (1950) & 62    & 13    & 1     & 2     & 1     & 1 \\
    \midrule
    Indoor1 (1890) & 87    & 13    & 3     & 2     & 1     & 1 \\
    \midrule
    Indoor2 (1530) & 61    & 13    & 1     & 1     & 2     & 3 \\
    \bottomrule
    \end{tabular}%
  \label{tab:image}%
\end{table*}%

\begin{table*}[t]
  \centering
  \caption{{\rm Number of time steps for different objects. The number in the parentheses is the groundtruth focus position.}}
 \begin{tabular}{ccccccc}
    \toprule
    Object & Rule-based & Fibonacci & Proposed-1100 & Proposed-1400 & Proposed-1700 & Proposed-2000 \\
    \midrule
    Toy1 (1830) & 46    & 13    & 2     & 2     & 1     & 2 \\
    \midrule
    Box1 (1360) & 47    & 13    & 1     & 2     & 1     & 1 \\
    \midrule
    Box2 (1410) & 51    & 13    & 1     & 0     & 3     & 1 \\
    \midrule
    Cup (1580) & 64    & 13    & 2     & 1     & 2     & 1 \\
    \midrule
    Bag (1240) & 39    & 13    & 1     & 2     & 1     & 5 \\
    \midrule
    Plant (1560) & 62    & 13    & 1     & 1     & 2     & 3 \\
    \midrule
    Book (1260) & 40    & 13    & 1     & 2     & 1     & 2 \\
    \midrule
    3D Printing (1300) & 43    & 13    & 1     & 1     & 1     & 2 \\
    \bottomrule
    \end{tabular}%
  \label{tab:table_object}%
\end{table*}%


\subsection{All-in-focus Imaging}
\subsubsection{System Configuration}
Because limited I/O speed and lag in focus adjustment of EVETAR module prevent it from fully demonstrating the performance of the focus strategy, we use CA378-AOIS camera module (4.52mm, F/2) occupied with Voice-Coil Motor (henceforth referred to as the VCM camera) by CenturyArks Co., Ltd. that achieves fast focus control. 

The camera was calibrated every 25 steps in the interval $(1000, 450)$, which corresponded to approximately $100$ millimeters to $1000$ millimeters in object distance. We trained the step estimator using the same network configuration as in Section \ref{sec:af-pipeline} and converted it to the global step estimator.

DoF across the range was also calibrated subjectively and then mapped to $\sigma_Z$. Also the width of each bin of the histogram was adjusted to match the DoF. 

\subsubsection{Static Scenes}
First we demonstrate the performance of the proposed strategy in conventional all-in-focus imaging when the scene is static. 

\begin{figure*}[htbp]
\centering
  {\footnotesize
      \begin{minipage}[c]{1.0\linewidth}
    		\begin{minipage}[b]{.72\linewidth}
    			\begin{tabular}[c]{cccc}
    				\subfigure{
    					\stackunder[4pt]{\includegraphics[width=0.21\linewidth]{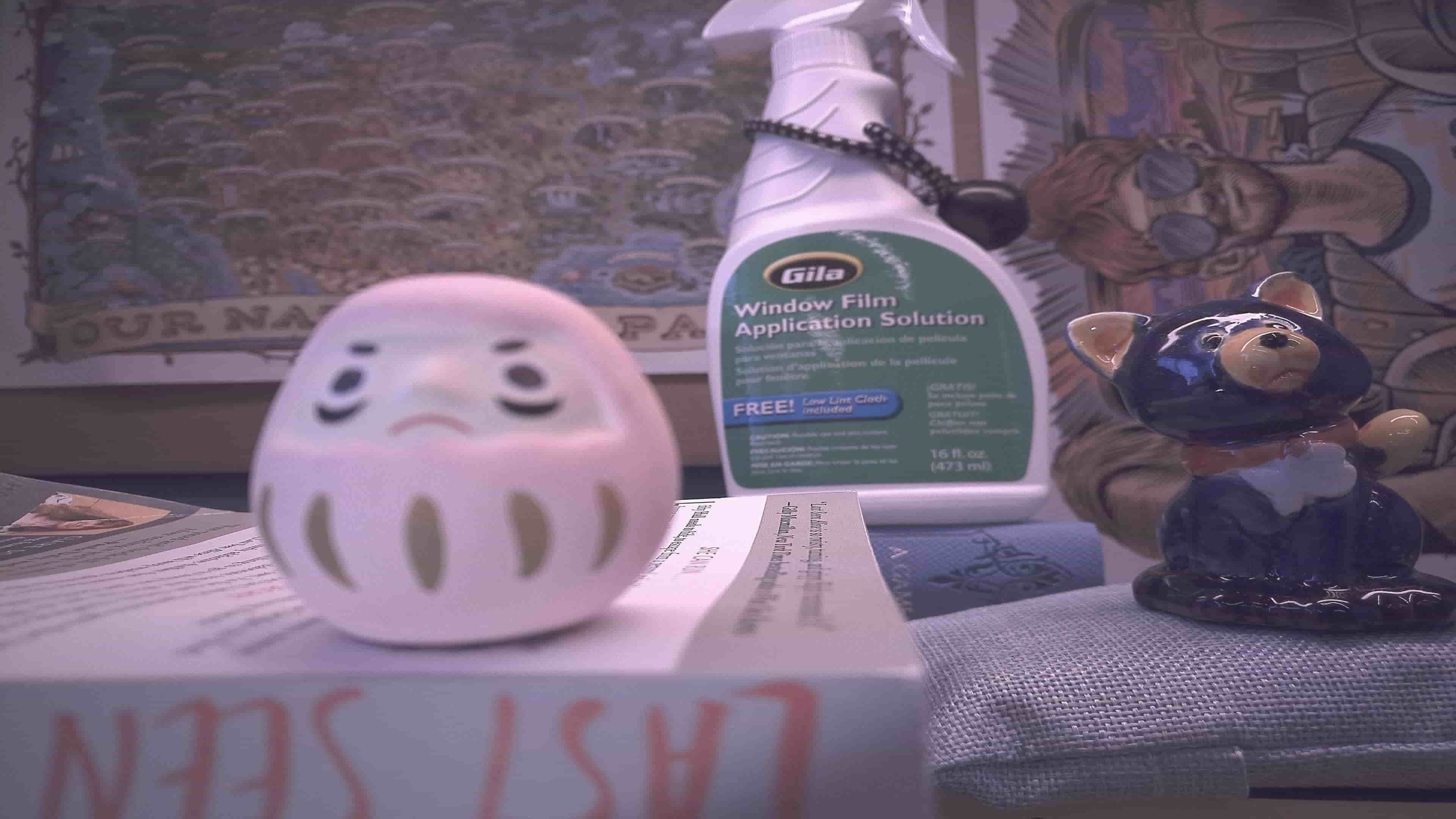}}{(a) $I^0$}}		&
    				\subfigure{
    					\stackunder[4pt]{\includegraphics[width=0.21\linewidth]{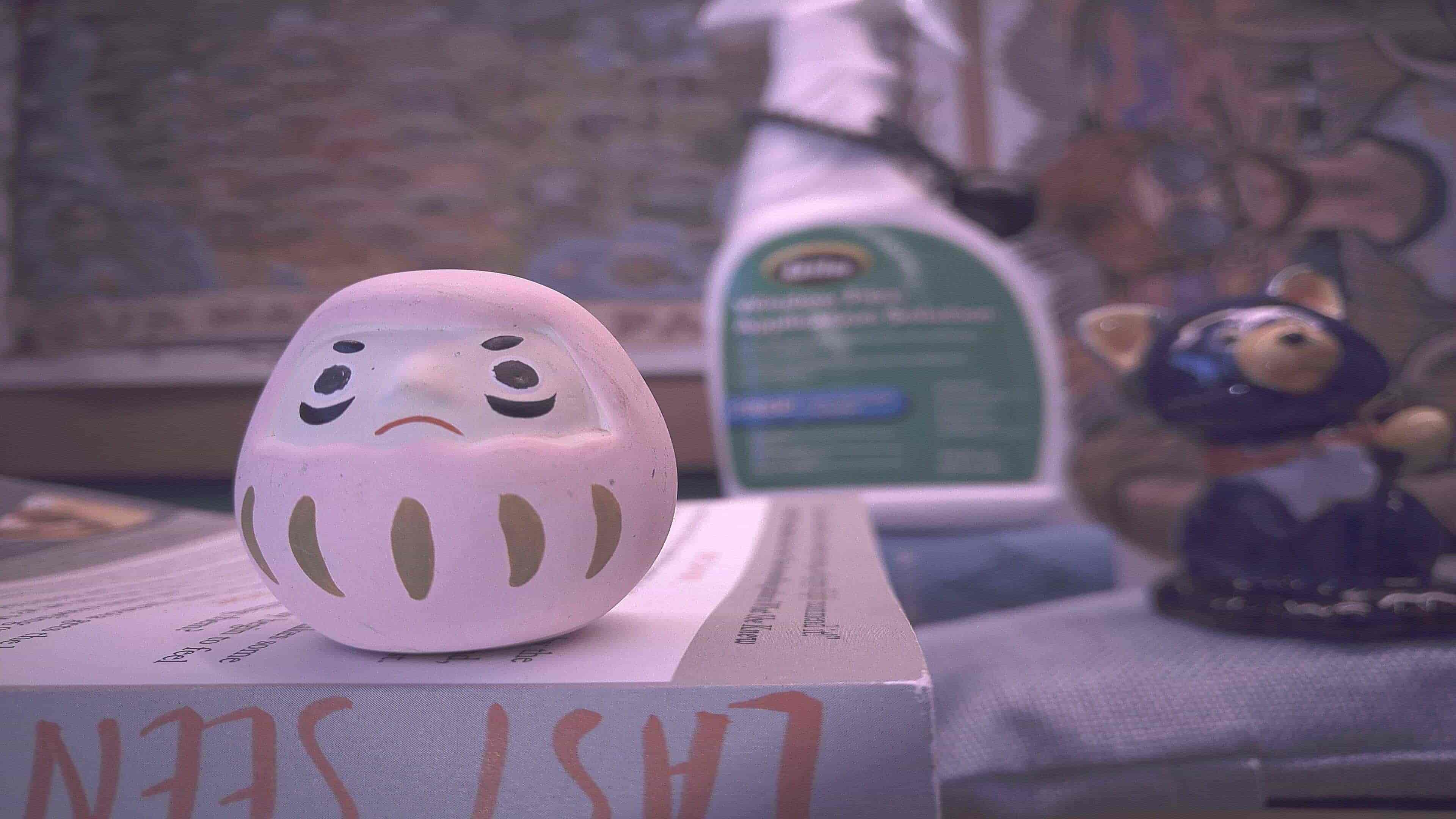}}{(b) $I^1$}}	 &
    				\subfigure{
    					\stackunder[4pt]{\includegraphics[width=0.21\linewidth]{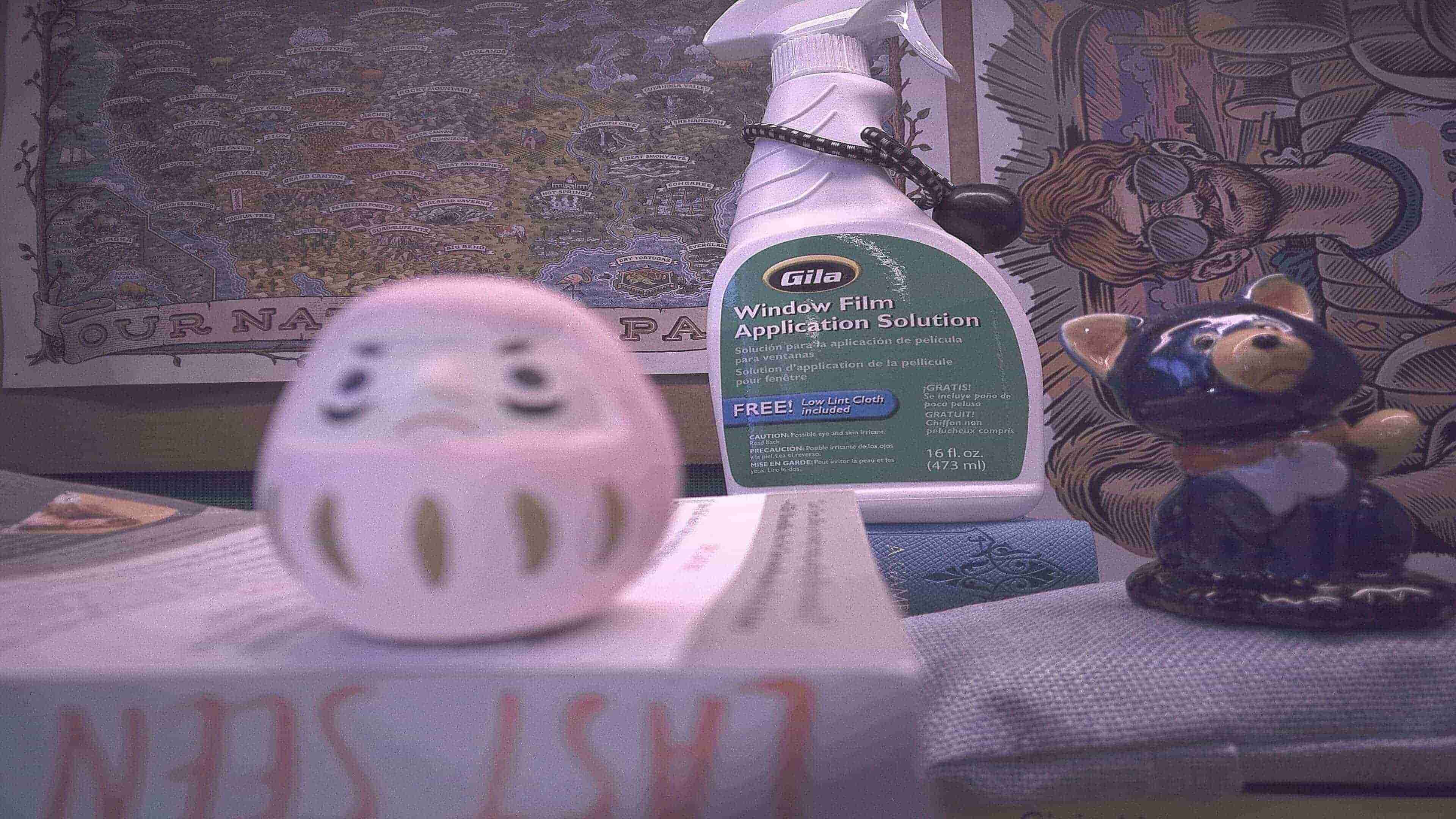}}{(c) $I^2$}}	 &
    				\subfigure{
    					\stackunder[4pt]{\includegraphics[width=0.21\linewidth]{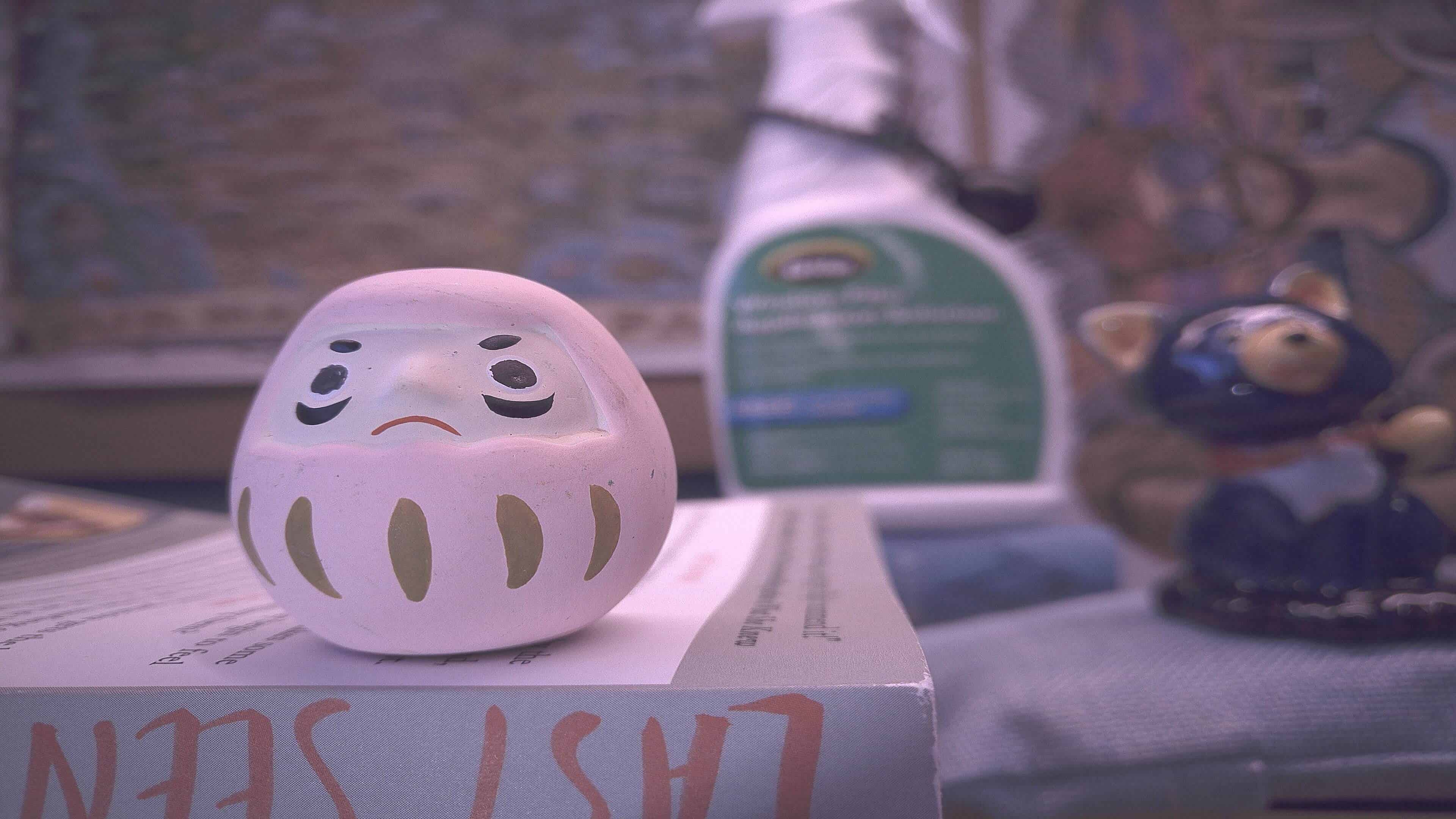}}{(d) $I^3$}}	\\
    				\subfigure{
    					\stackunder[4pt]{\fcolorbox{black}{black}{\includegraphics[width=0.21\linewidth]{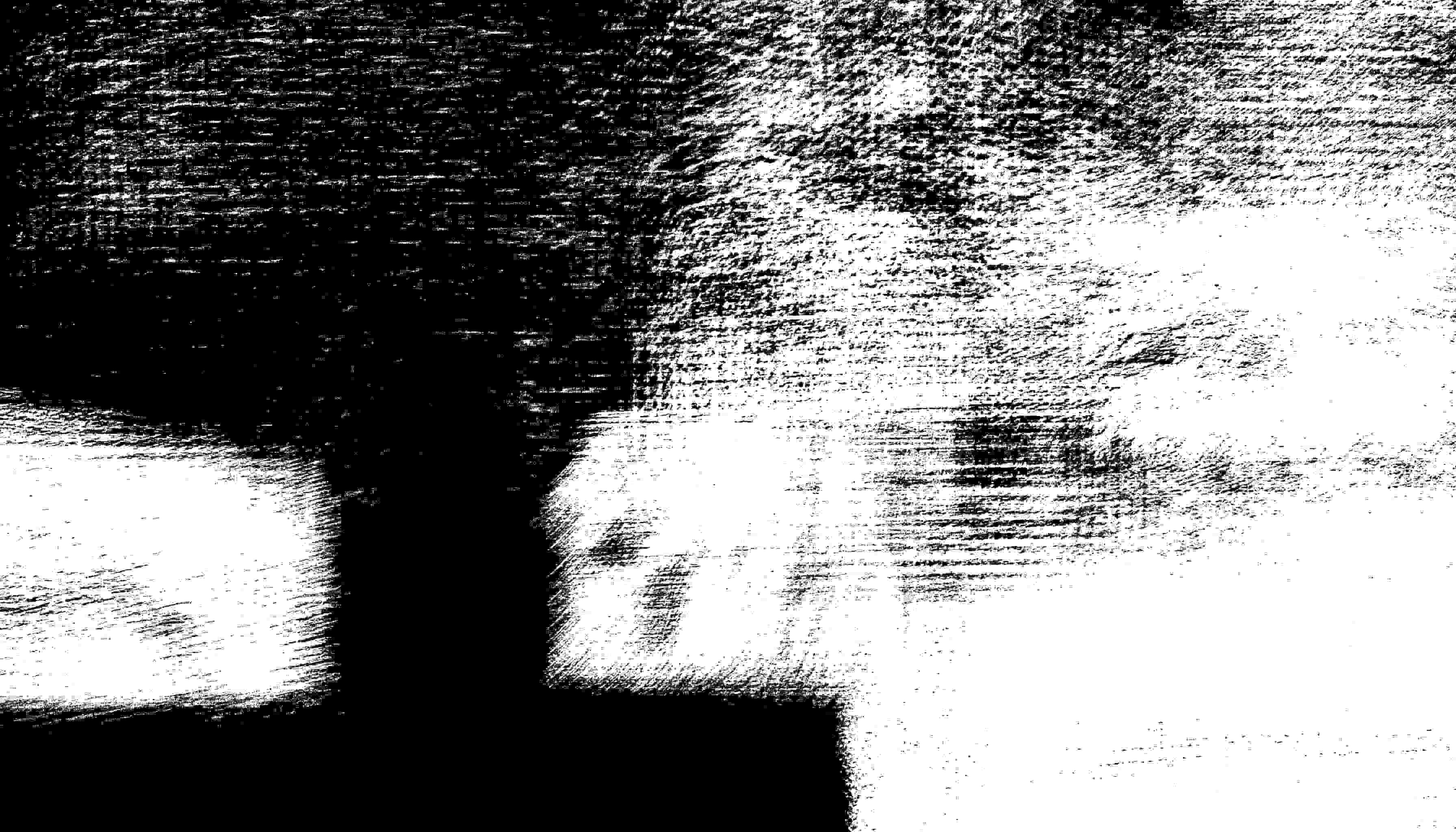}}}{(e) $U^0$}}		& 
    				\subfigure{
    					\stackunder[4pt]{\fcolorbox{black}{black}{\includegraphics[width=0.21\linewidth]{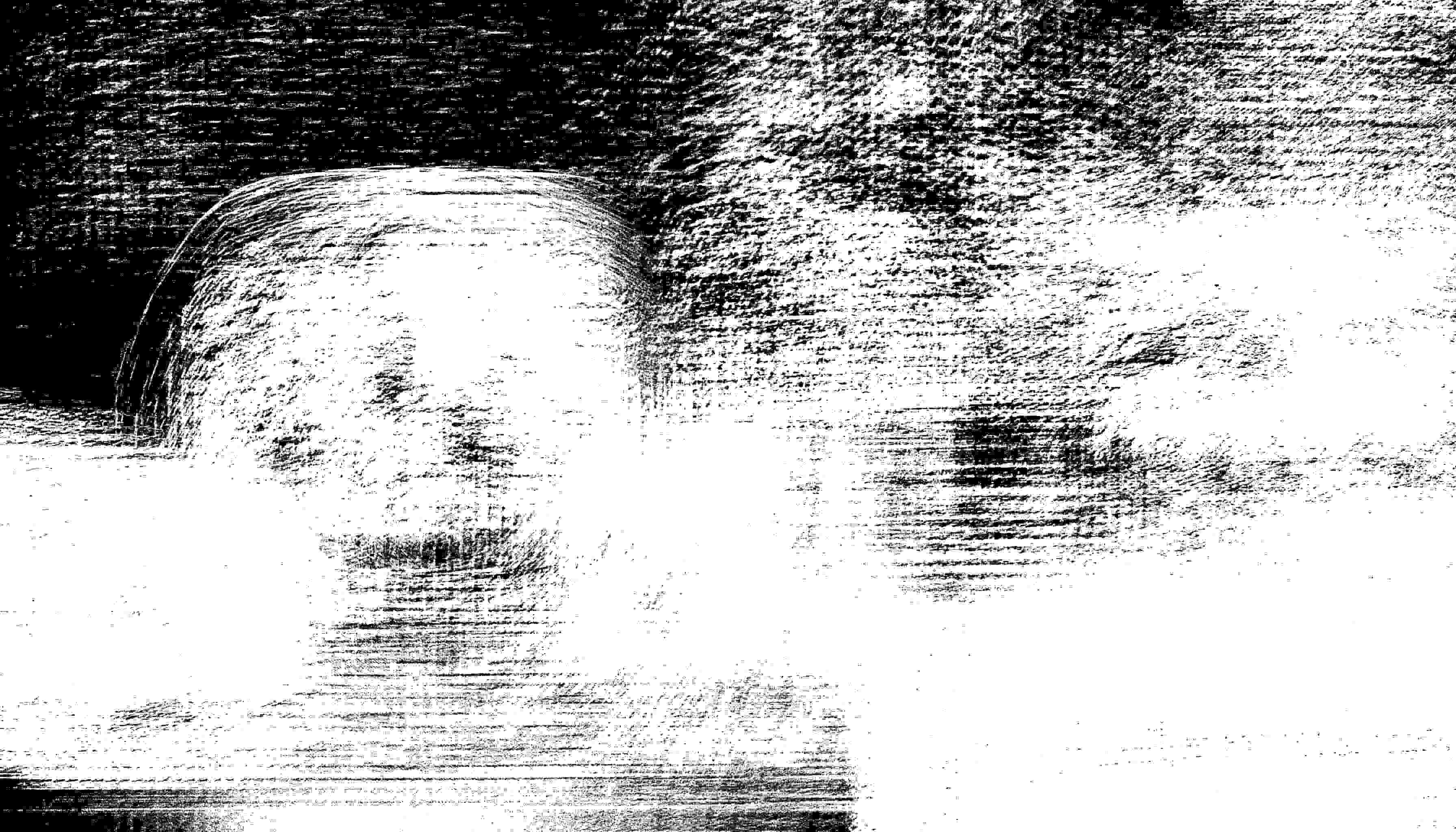}}}{(f) $U^1$}}	 & 
    				\subfigure{
    					\stackunder[4pt]{\fcolorbox{black}{black}{\includegraphics[width=0.21\linewidth]{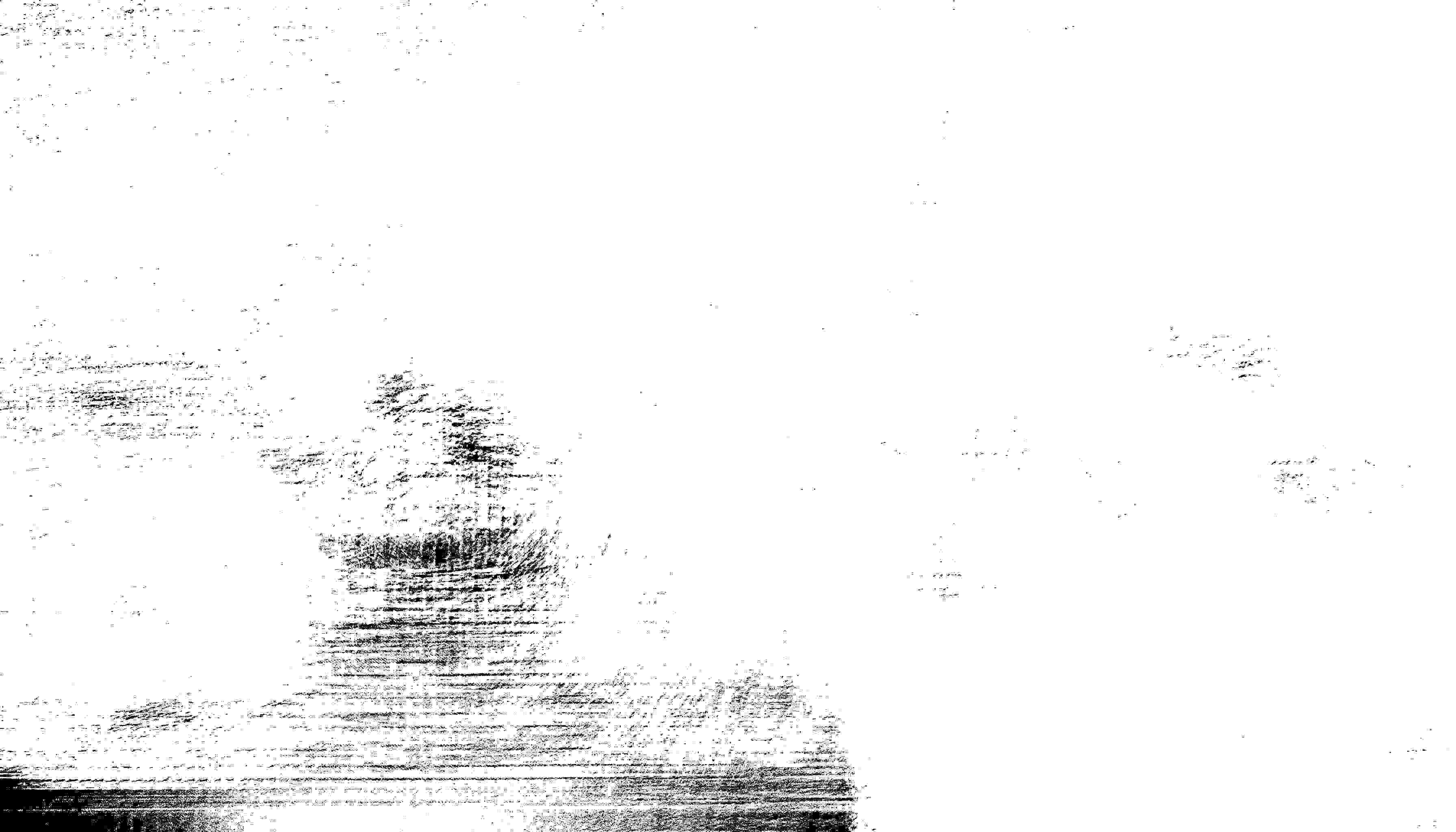}}}{(g) $U^2$}}	 &
    				\subfigure{
    					\stackunder[4pt]{\fcolorbox{black}{black}{\includegraphics[width=0.21\linewidth]{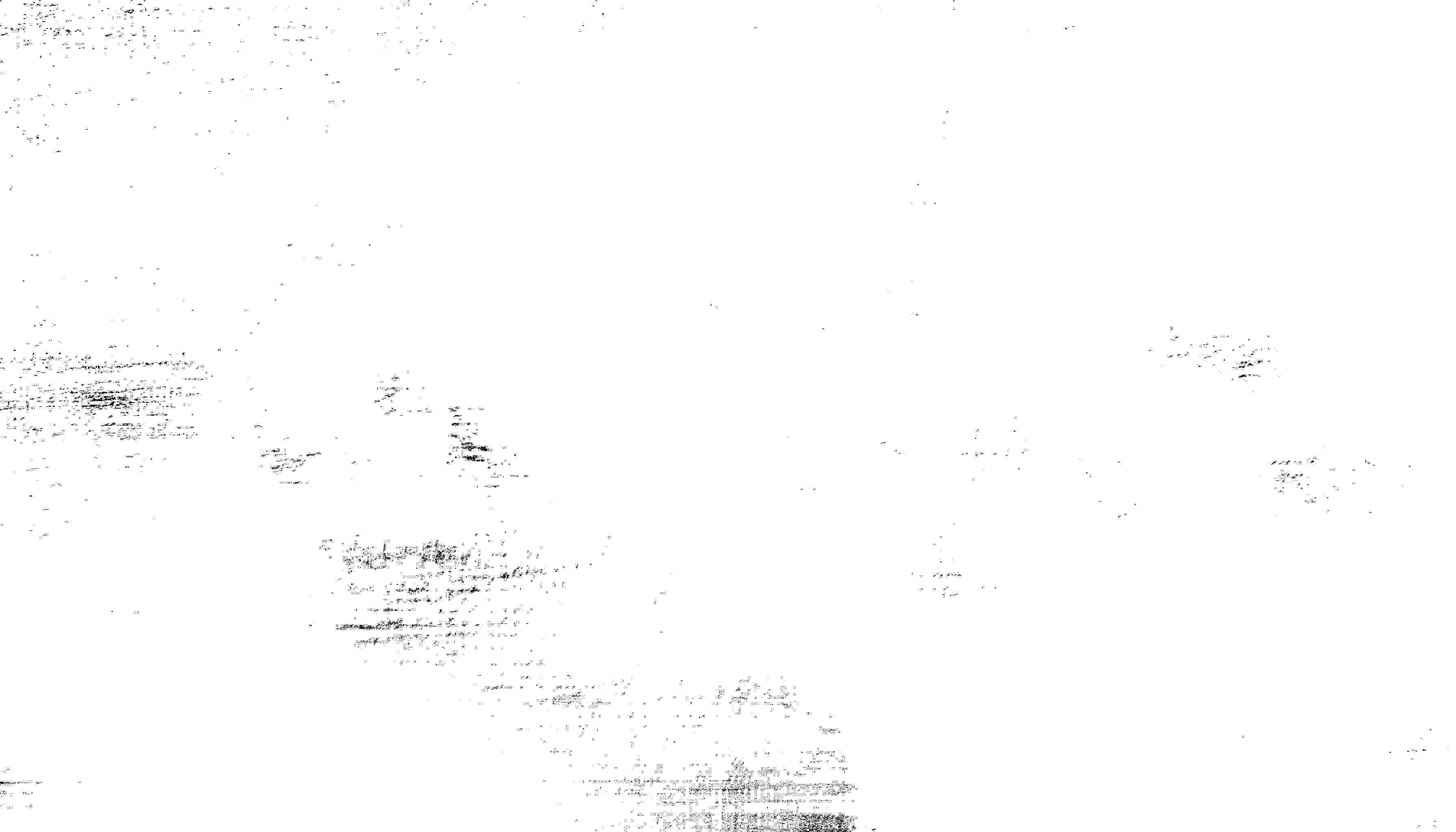}}}{(h) $U^3$}}	
    			\end{tabular}
    		\end{minipage}
    		\begin{minipage}[c]{0.28\linewidth}
    		    \centering
    			\vspace*{6px}
    			\subfigure{
    				\stackunder[5pt]{\includegraphics[width=1\linewidth]{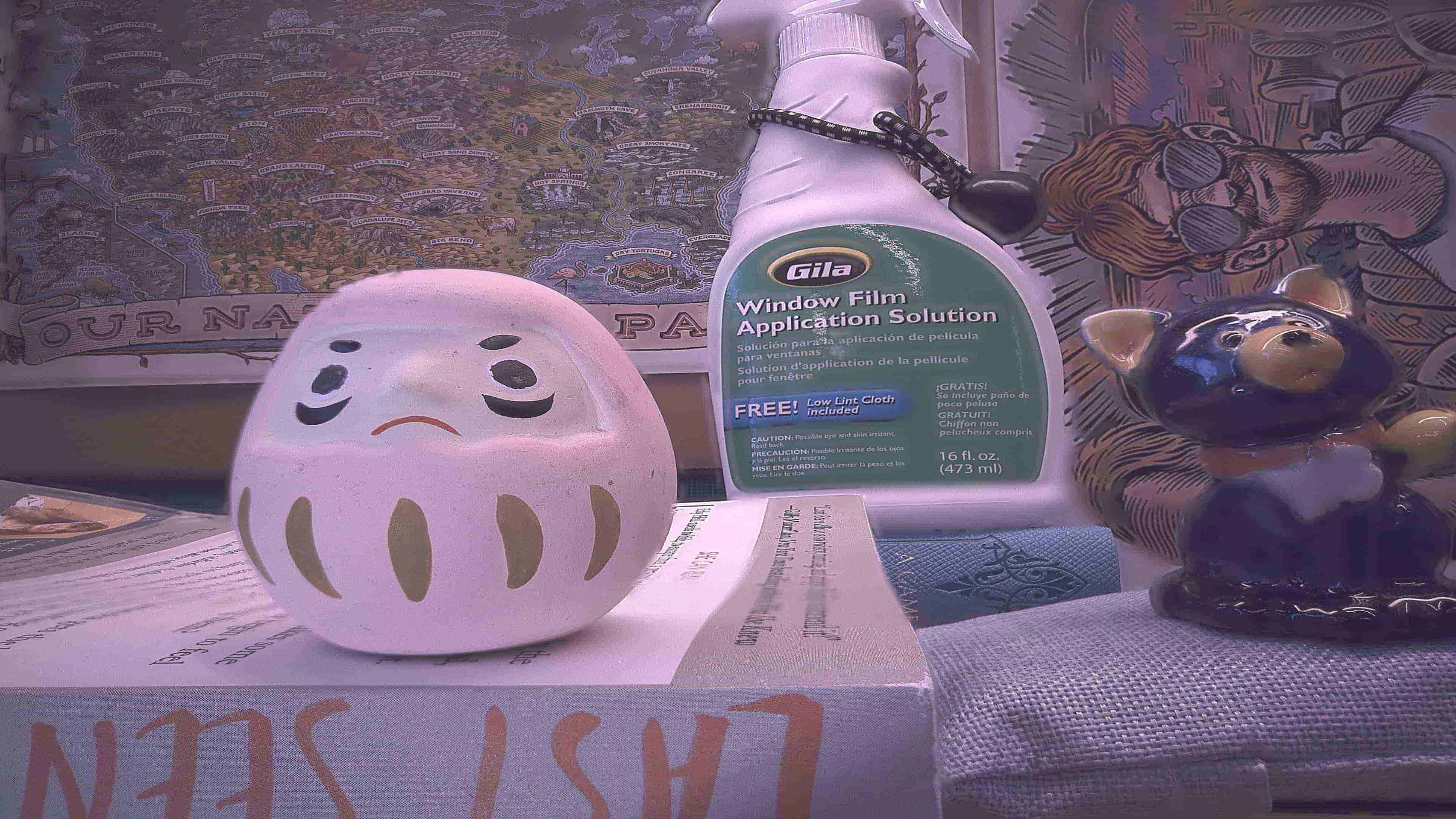}}{(I)}
    			}		
    		\end{minipage} 
    	\end{minipage}
    	\begin{minipage}[c]{1.0\linewidth}
    		\begin{minipage}[b]{.72\linewidth}
    			\begin{tabular}[c]{cccc}
    				\subfigure{
    					\stackunder[4pt]{\includegraphics[width=0.21\linewidth]{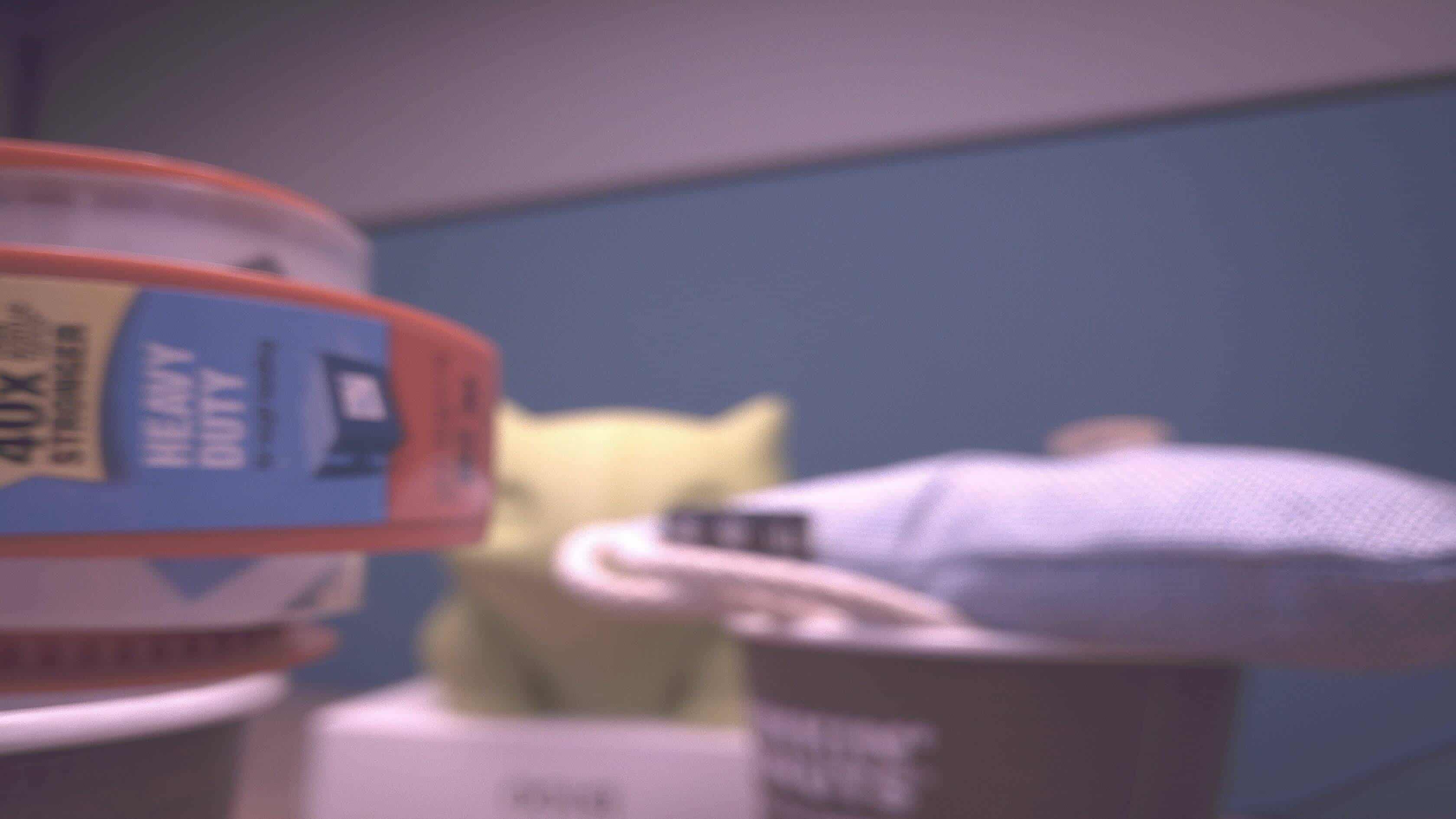}}{(a) $I^0$}}		&
    				\subfigure{
    					\stackunder[4pt]{\includegraphics[width=0.21\linewidth]{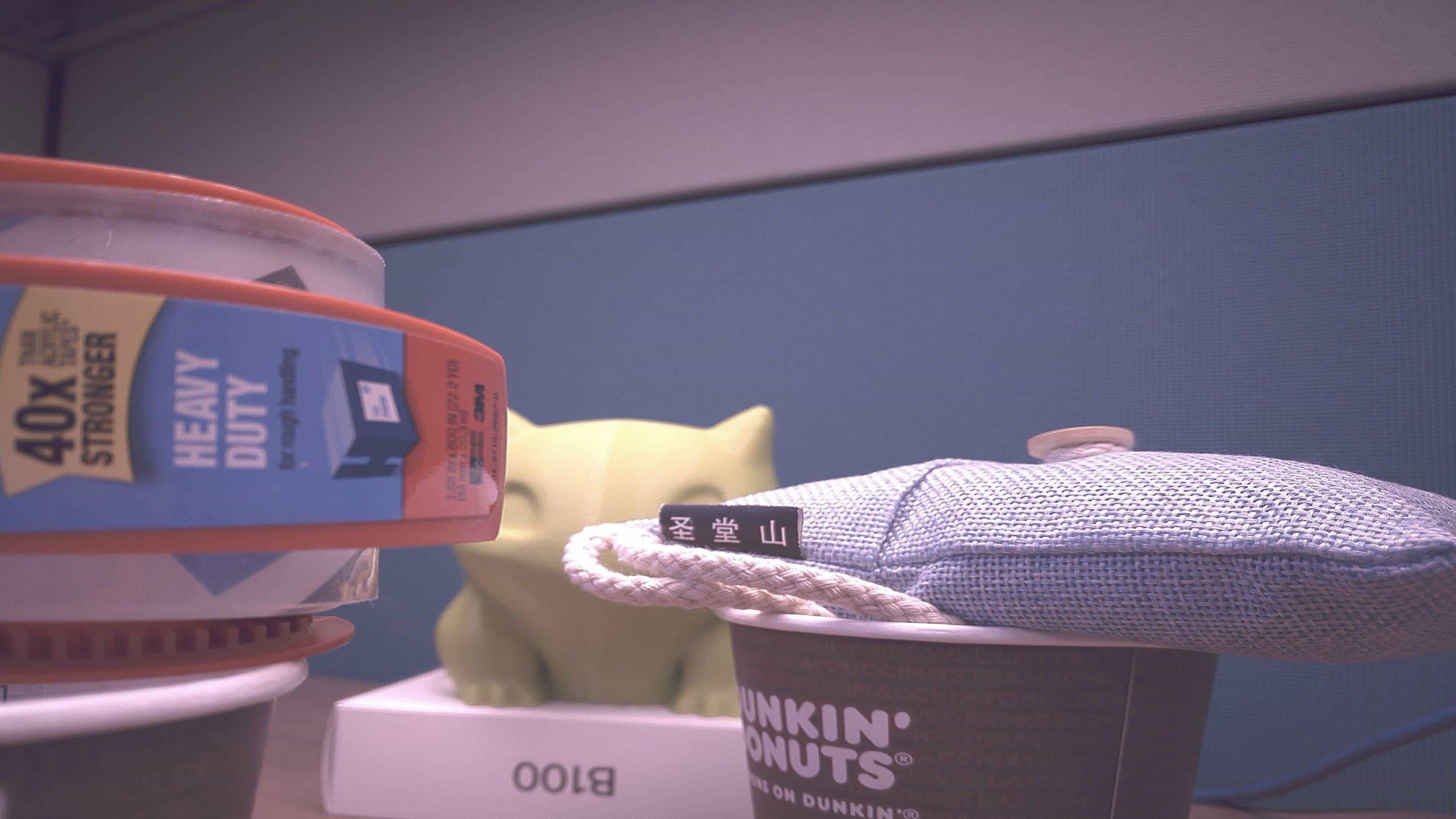}}{(b) $I^1$}}	 &
    				\subfigure{
    					\stackunder[4pt]{\includegraphics[width=0.21\linewidth]{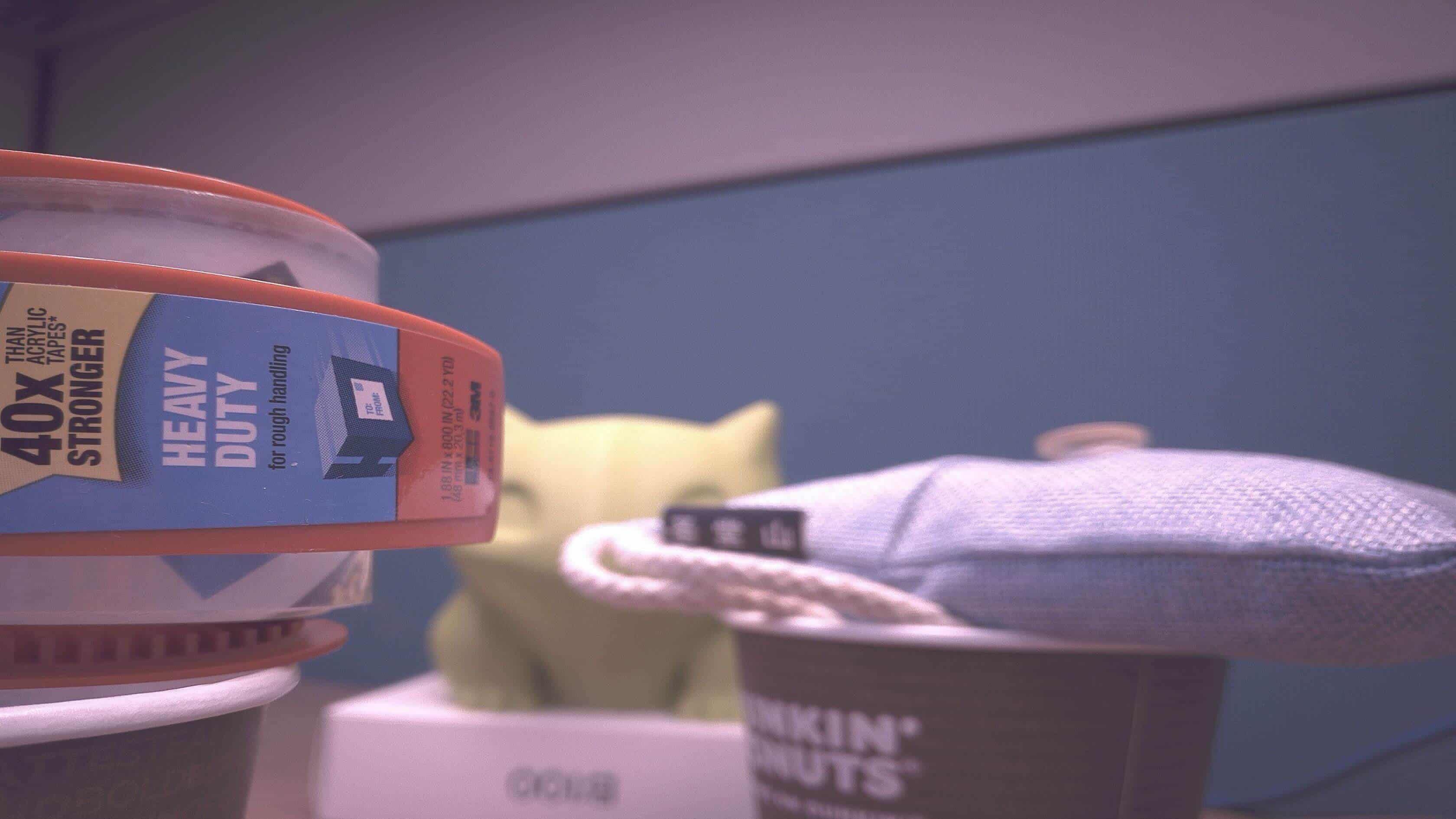}}{(c) $I^2$}}	 &
    				\subfigure{
    					\stackunder[4pt]{\includegraphics[width=0.21\linewidth]{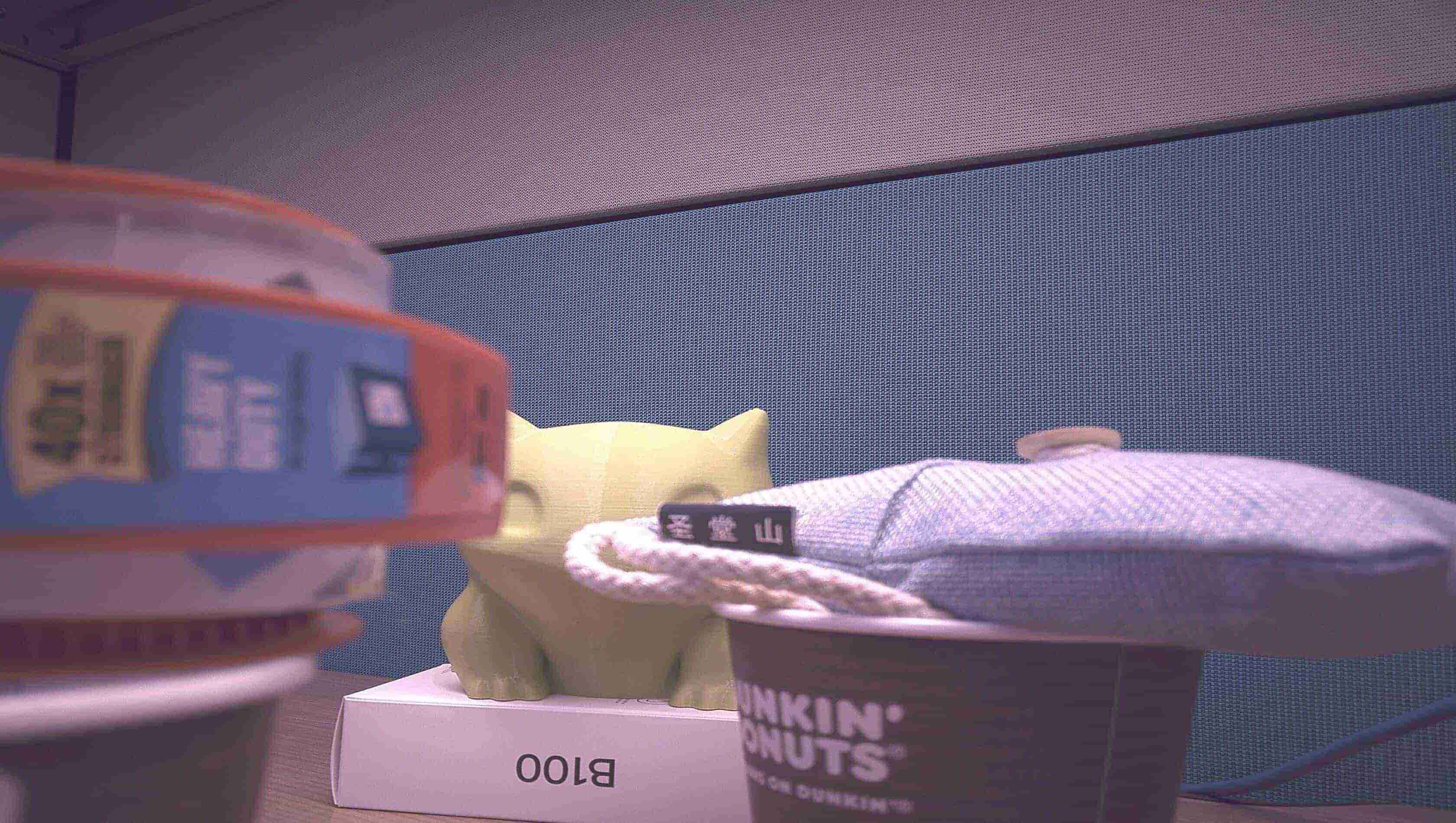}}{(d) $I^3$}}	\\
    				\subfigure{
    					\stackunder[4pt]{\fcolorbox{black}{black}{\includegraphics[width=0.21\linewidth]{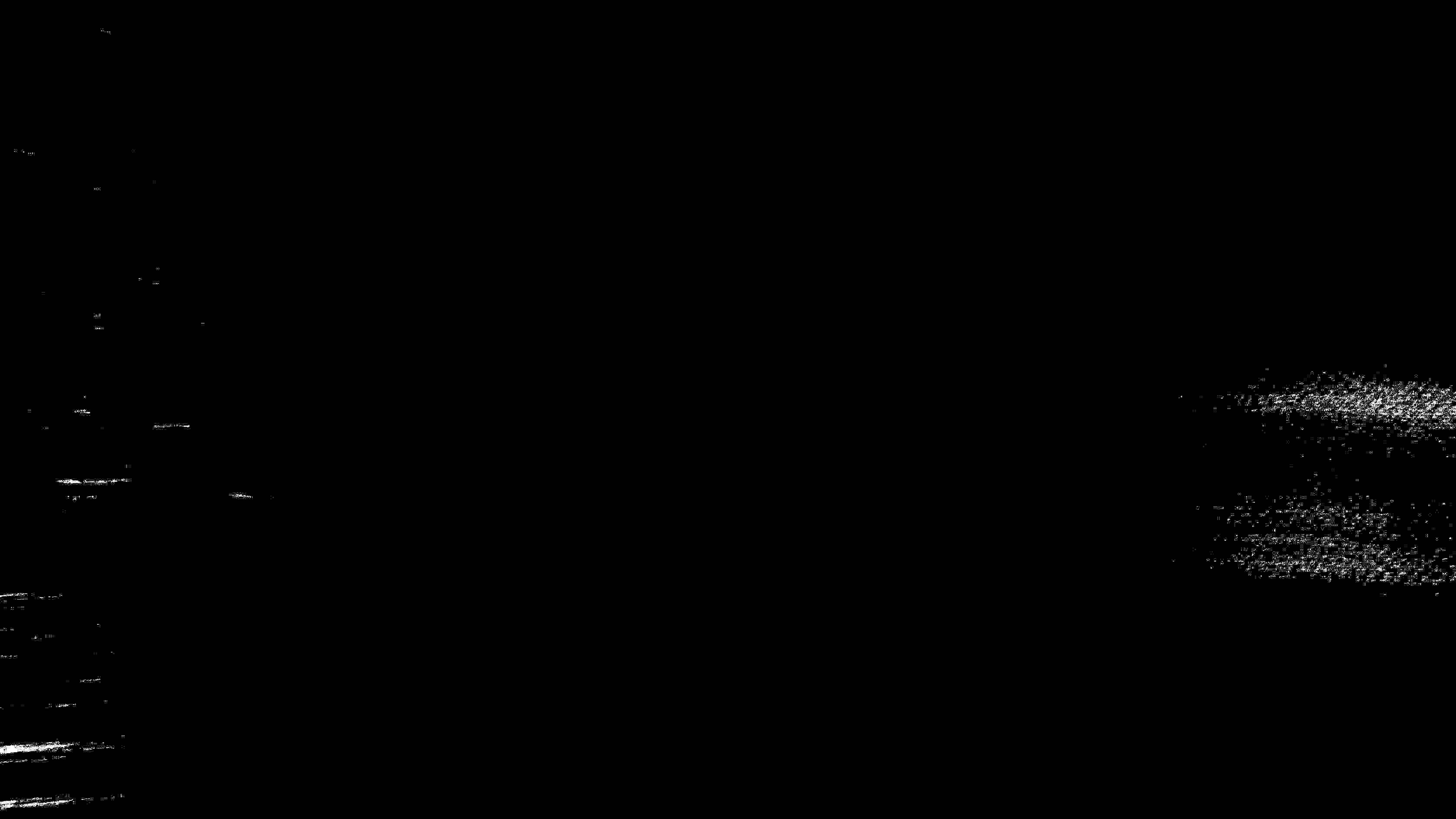}}}{(e) $U^0$}}		& 
    				\subfigure{
    					\stackunder[4pt]{\fcolorbox{black}{black}{\includegraphics[width=0.21\linewidth]{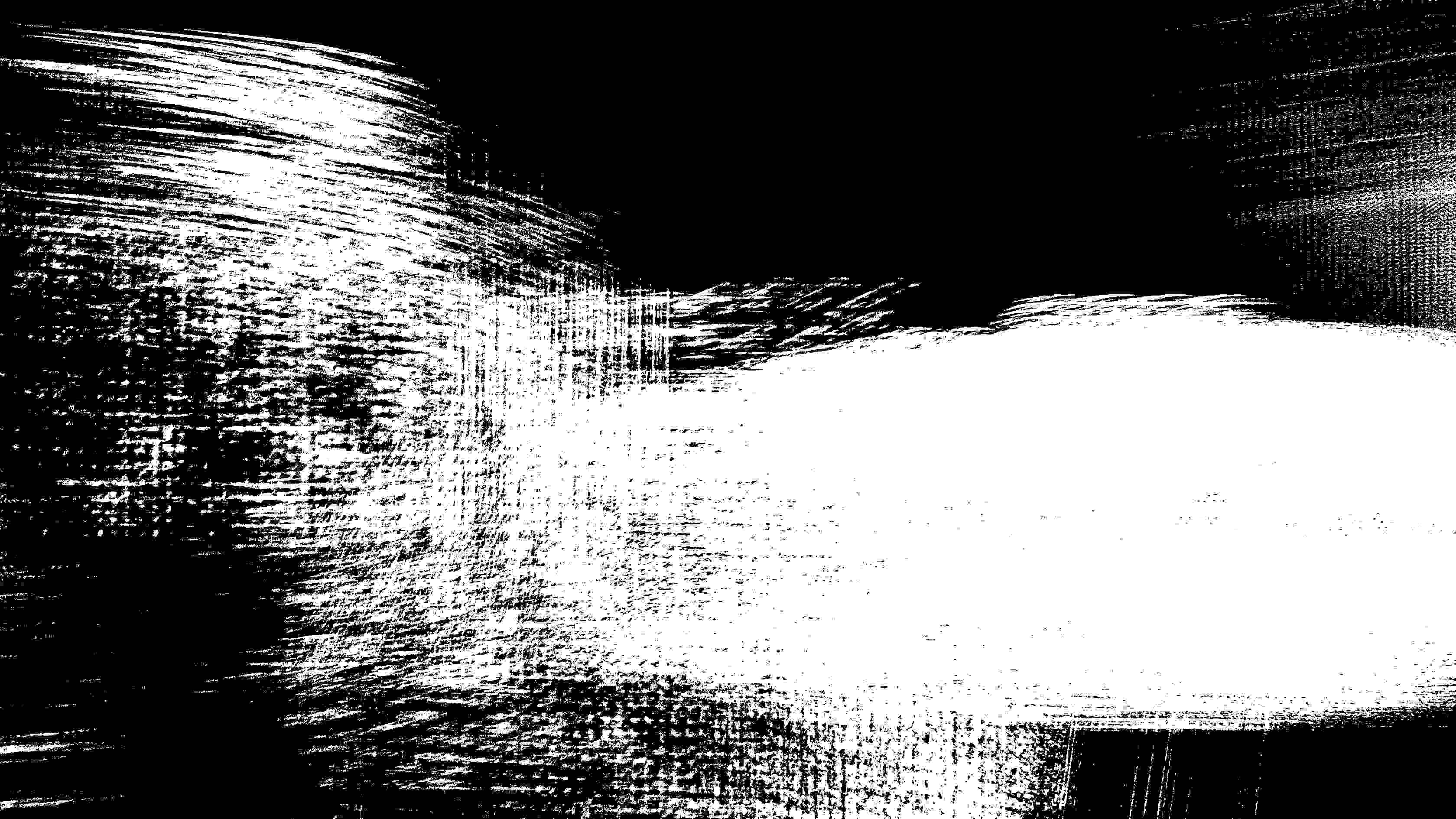}}}{(f) $U^1$}}	 & 
    				\subfigure{
    					\stackunder[4pt]{\fcolorbox{black}{black}{\includegraphics[width=0.21\linewidth]{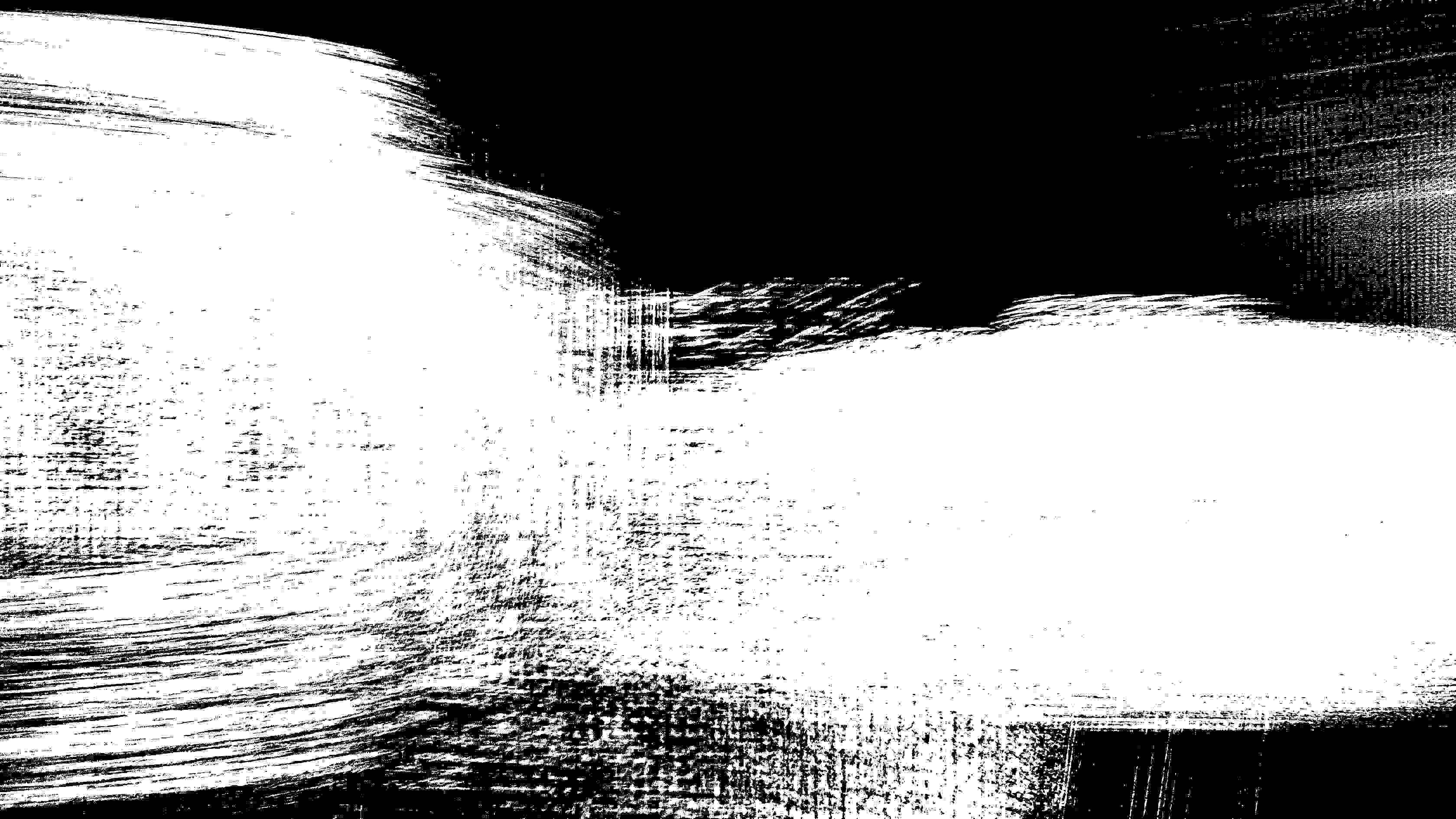}}}{(g) $U^2$}}	 &
    				\subfigure{
    					\stackunder[4pt]{\fcolorbox{black}{black}{\includegraphics[width=0.21\linewidth]{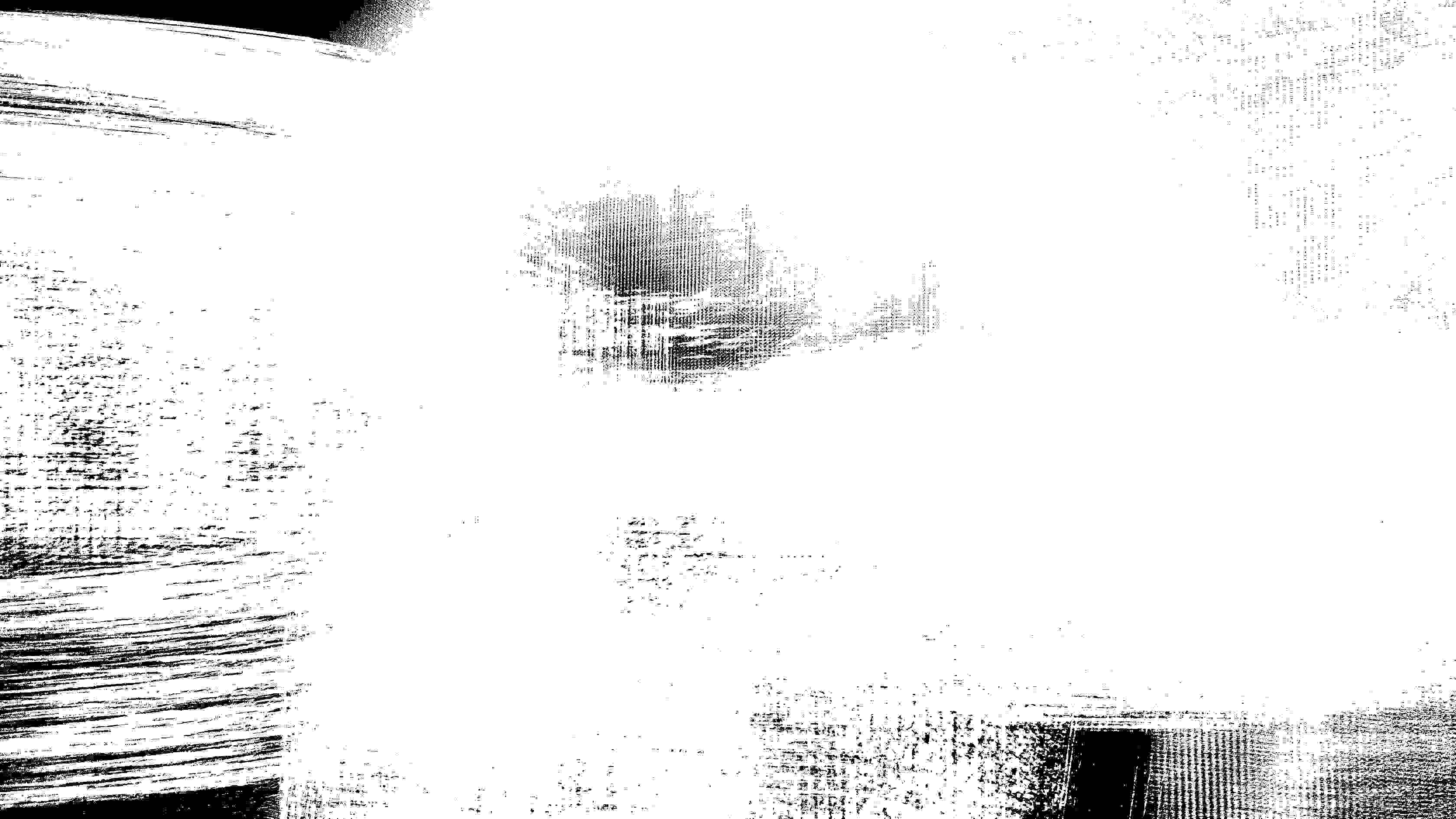}}}{(h) $U^3$}}	
    			\end{tabular}
    		\end{minipage}
    		\begin{minipage}[c]{0.28\linewidth}
    		    \centering
    			\vspace*{6px}
    			\subfigure{
    				\stackunder[5pt]{\includegraphics[width=1\linewidth]{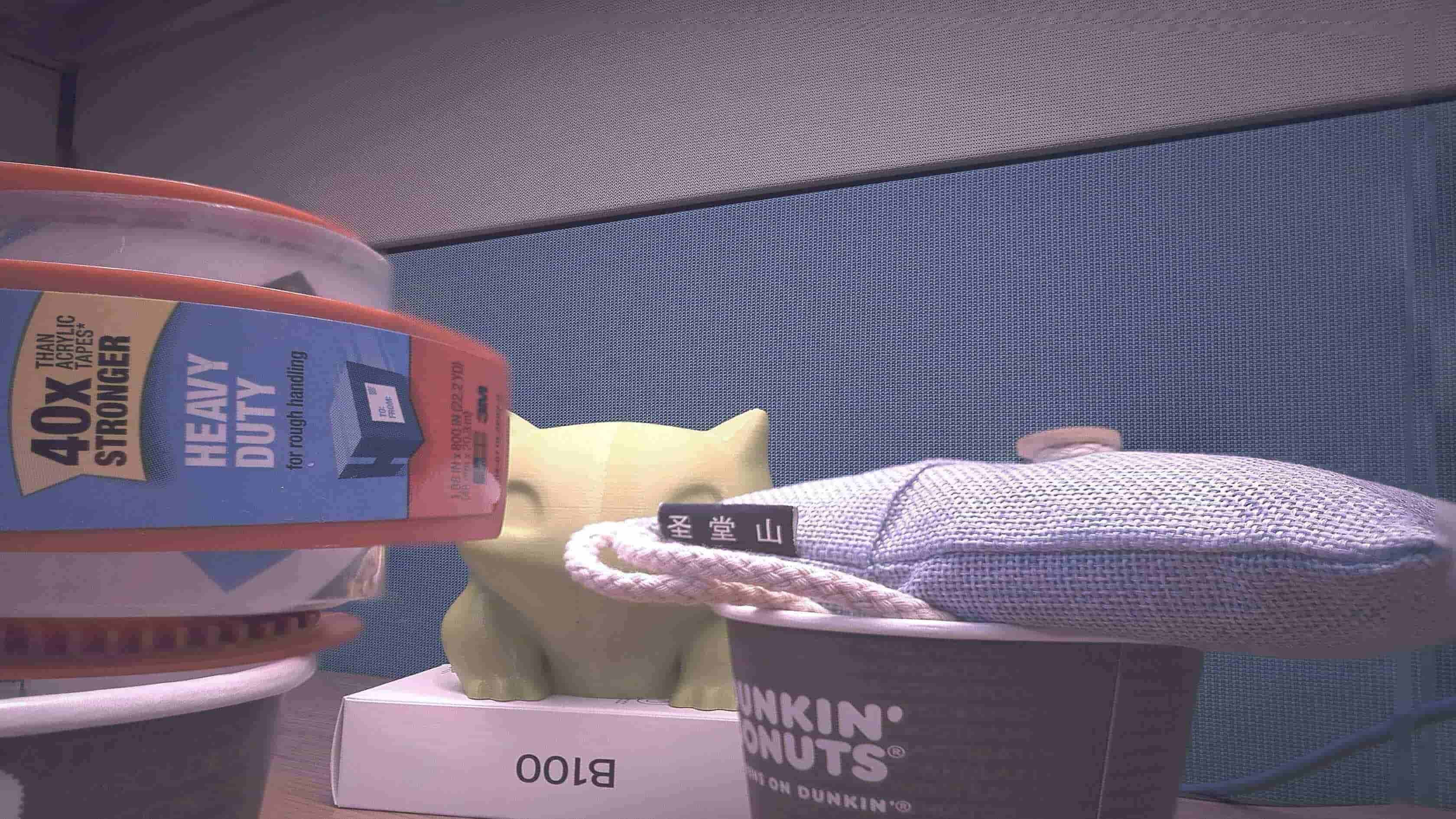}}{(II)}
    			}		
    		\end{minipage} 
    	\end{minipage}
    \caption{Two all-in-focus images generated using the proposed focus strategy. In each example, (a)-(d) are the captured frames, and (e)-(h) are the activation matrices. (I) and (II) are the fused all-in-focus images by \cite{QIU201935}. Note in the first example the title of the foremost book was not fully in-focus in (b), thus (d) was captured.\label{fig: static fs results}}
	}
\end{figure*}

We shot two all-in-focus images. Each time several objects were randomly placed in front of the camera. The captured frames along with the corresponding activation matrices are shown in Figure \ref{fig: static fs results}.

To fuse the frames, we applied the method proposed in \cite{QIU201935} (codes available at \url{https://github.com/bitname/Multi-focus-image-fusion-GFDF}). The ratio factor $R$ was set to 0.001. To address the focus breathing problem, the frames were first aligned using SURF\cite{bay2006surf}.

Compared with existing scene-adaptive methods~\cite{vaquero2011generalized,choi2017improved,li2018scene}, the proposed method requires neither a full range scanning nor a refinement step. Only informative frames are captured, and these positions are visited only once during the capture process. This results in much fewer captured frames and runtime. 

It should be emphasized that the proposed strategy is flexible in detailed implementation. To avoid expensive computation, the global step estimator can sparsify the patches that need to be processed by changing the stride and dilation in the convolutional layers. We can also update $P$ by considering all the previous $P$s and $Z$s to improve the robustness.

\fboxsep=0.1mm
\fboxrule=0pt

\subsubsection{Dynamic Scenes}
In dynamic scenes, all-in-focus imaging aims at producing videos in full focus. We test this by capturing a sequence of frames and outputting the fused all-in-focus frames.

\begin{figure}
	\centering
	\subfigure{\includegraphics[width=0.96\linewidth]{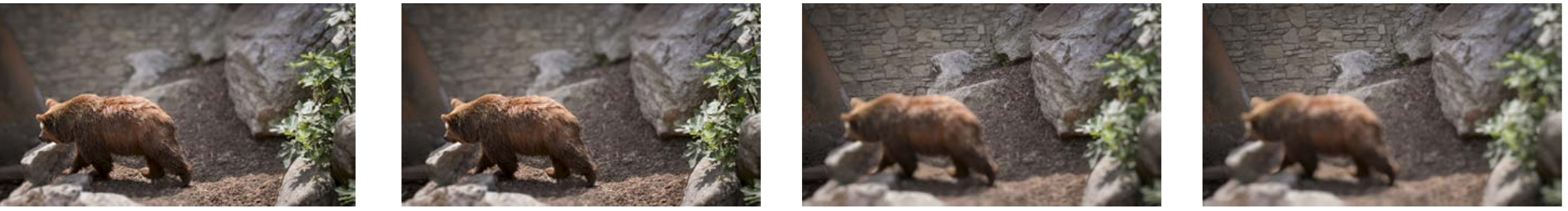}}
	\caption{Simulated defocused frames from DAVIS dataset.}
	\label{fig: dfs training data}
\end{figure}

We applied AWnet~\cite{cheng2019dual} to fuse frames. AWnet transfers pixels from a reference image to the target image. Here in our problem, the fused frame from the previous time step is the reference image, and the captured frame is the target image. To train the AWnet, we simulated defocused frames from DAVIS dataset\cite{Perazzi2016}. We assigned the estimated depth using algorithm \cite{li2019learning} to the video clips, and then use calibrated camera defocus model to blur the frames. Figure \ref{fig: dfs training data} shows some of our training samples.

\begin{figure*}[htbp]
\centering
  {\footnotesize
      \begin{minipage}[c]{1.0\linewidth}
    			\begin{tabular}{cccccc}
    				\subfigure{
    					\stackunder[4pt]{\includegraphics[width=0.14\linewidth]{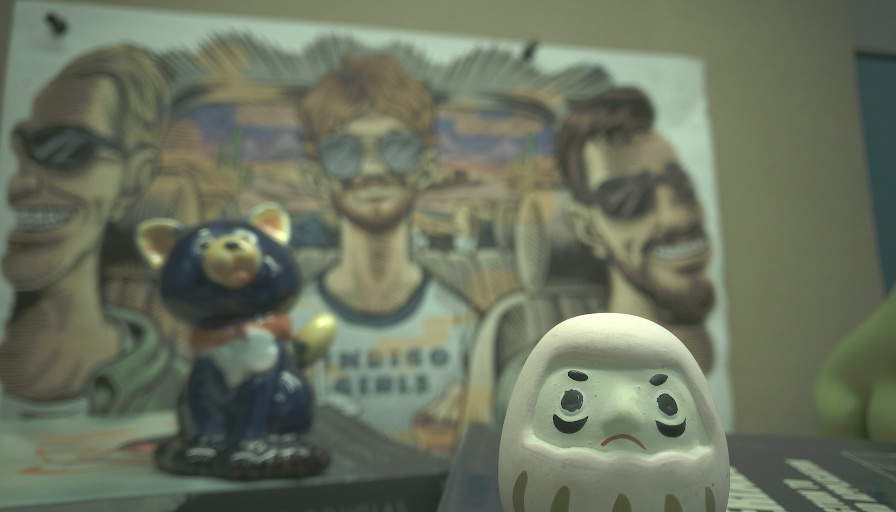}}{$I^0$}}		&
    				\subfigure{
    					\stackunder[4pt]{\includegraphics[width=0.14\linewidth]{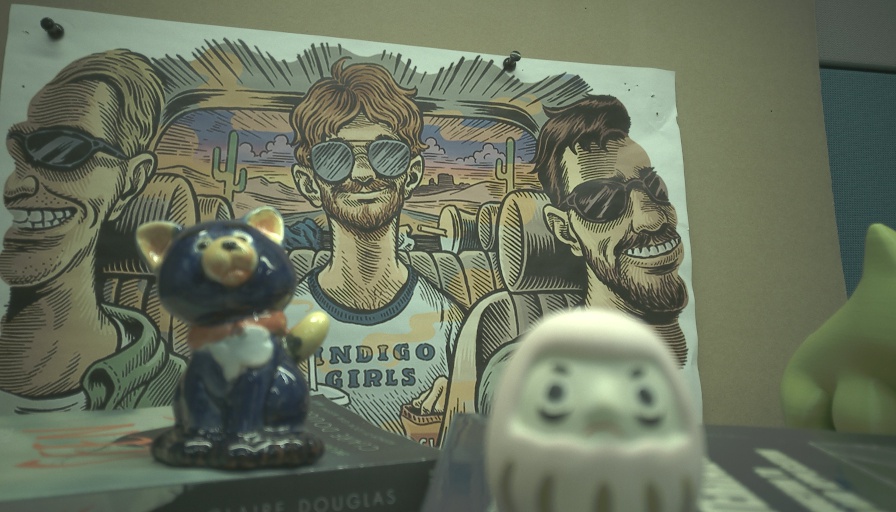}}{$I^1$}}	 &
    				\subfigure{
    					\stackunder[4pt]{\includegraphics[width=0.14\linewidth]{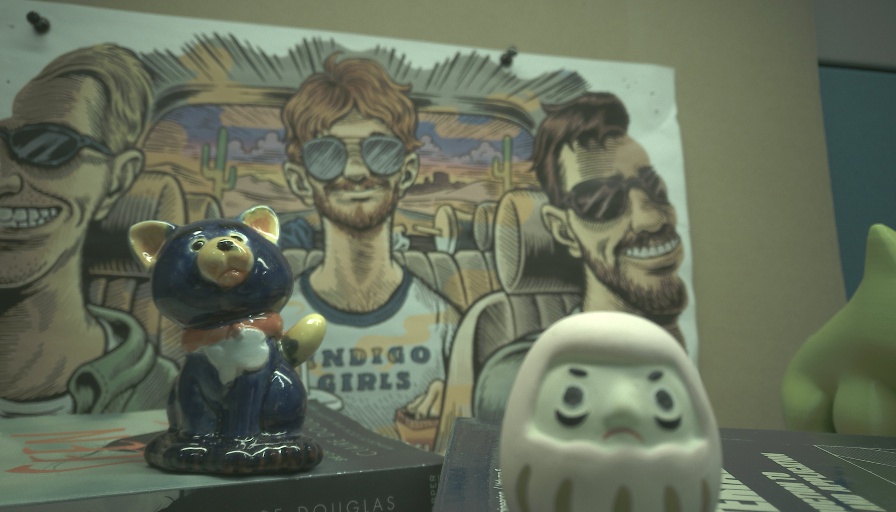}}{$I^2$}}	 &
    				\subfigure{
    					\stackunder[4pt]{\includegraphics[width=0.14\linewidth]{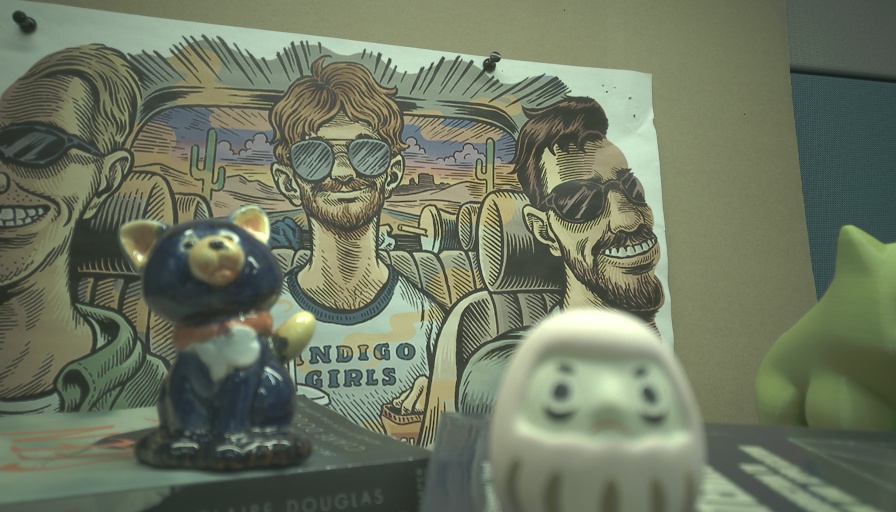}}{$I^3$}}	 &
    				\subfigure{
    					\stackunder[4pt]{\includegraphics[width=0.14\linewidth]{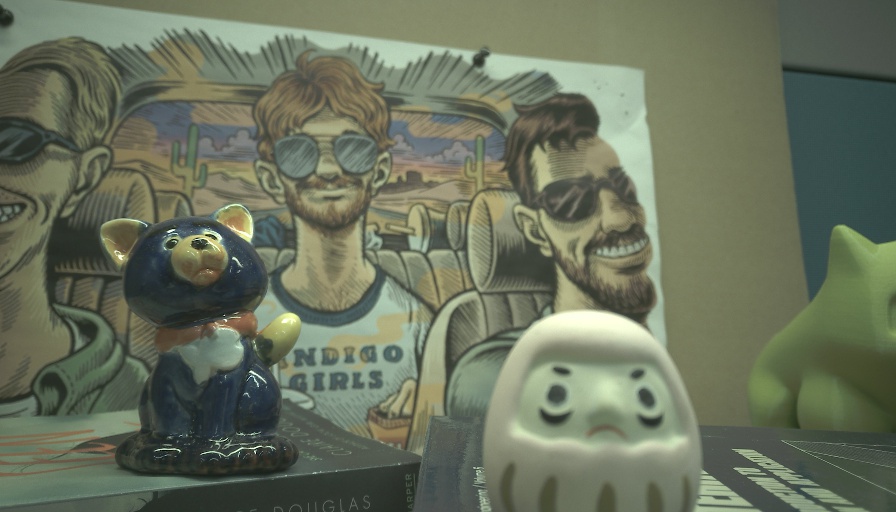}}{$I^4$}}	 &
    				\subfigure{
    					\stackunder[4pt]{\includegraphics[width=0.14\linewidth]{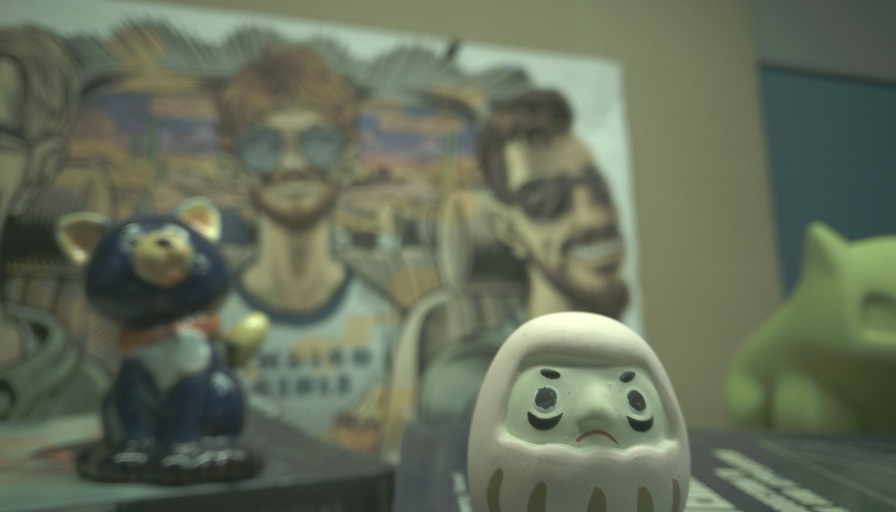}}{$I^5$}}	\\
    				\subfigure{
    					\stackunder[4pt]{\includegraphics[width=0.14\linewidth]{dfs_6frames/11717_s650_00.jpg}}{$J^0$}}		&
    				\subfigure{
    					\stackunder[4pt]{\includegraphics[width=0.14\linewidth]{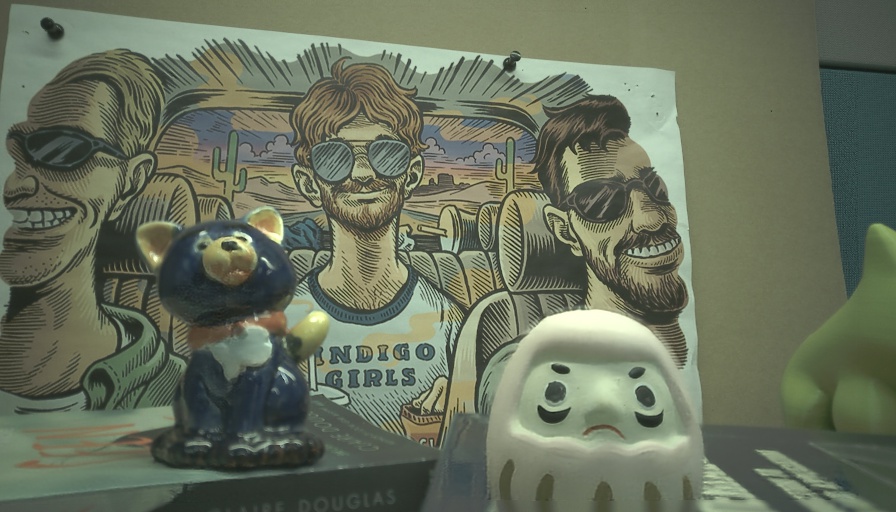}}{$J^1$}}	 &
    				\subfigure{
    					\stackunder[4pt]{\includegraphics[width=0.14\linewidth]{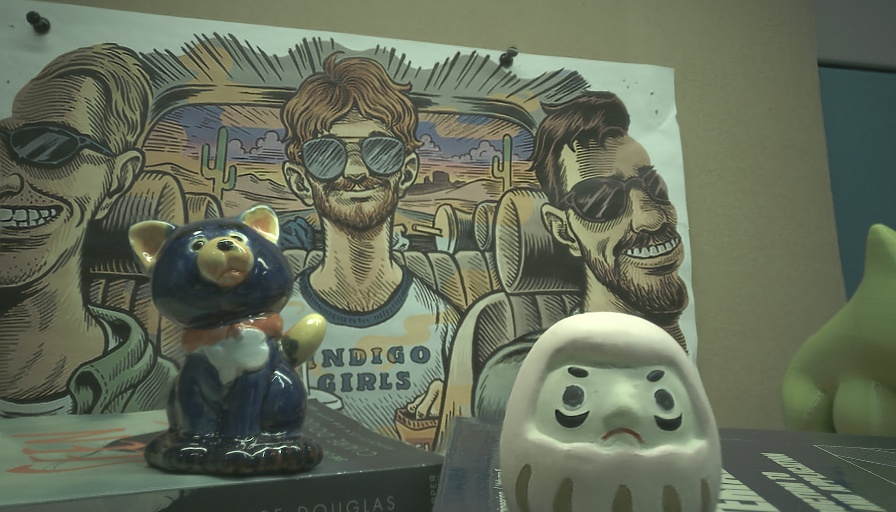}}{$J^2$}}	 &
    				\subfigure{
    					\stackunder[4pt]{\includegraphics[width=0.14\linewidth]{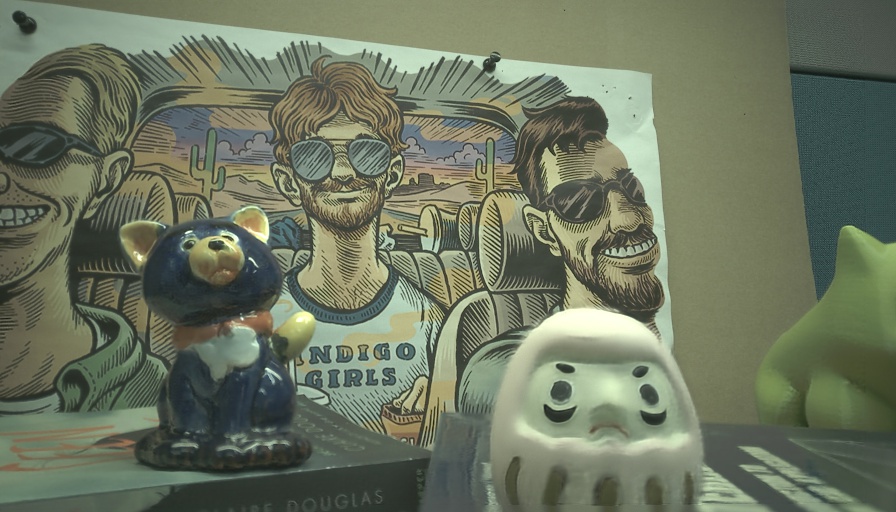}}{$J^3$}}	 &
    				\subfigure{
    					\stackunder[4pt]{\includegraphics[width=0.14\linewidth]{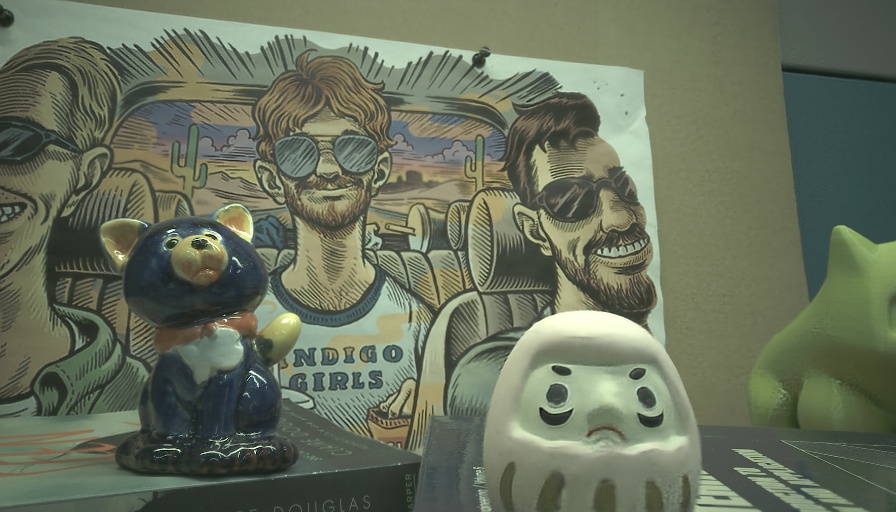}}{$J^4$}}	 &
    				\subfigure{
    					\stackunder[4pt]{\includegraphics[width=0.14\linewidth]{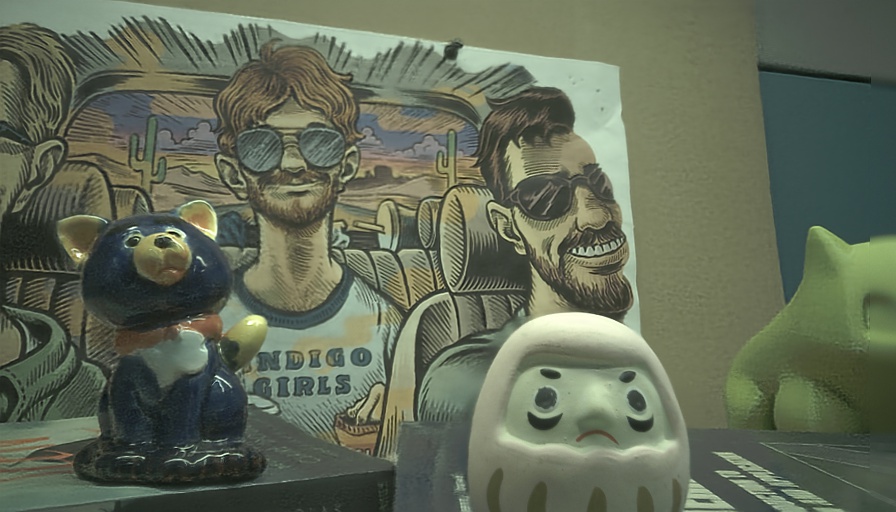}}{$J^5$}}
    			\end{tabular}
    			\begin{tabular}{cccccc}
    				\subfigure{
    					\stackunder[4pt]{\includegraphics[width=0.14\linewidth]{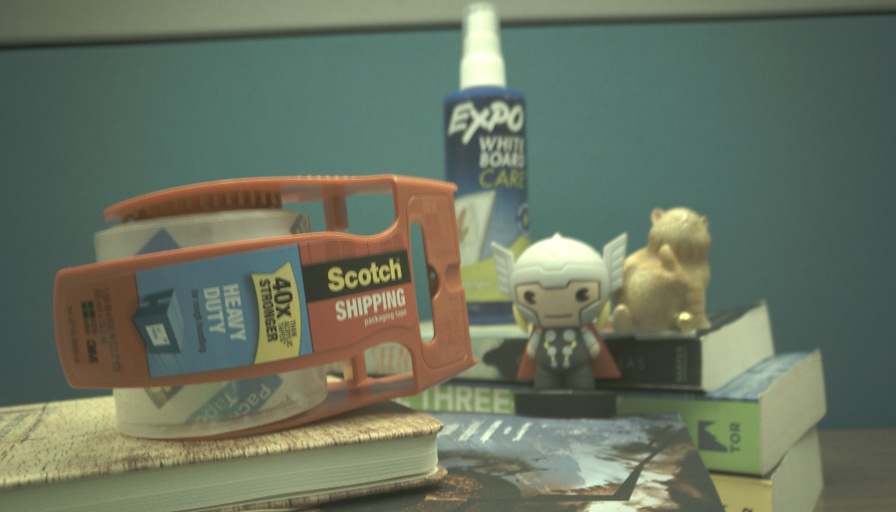}}{$I^0$}}		&
    				\subfigure{
    					\stackunder[4pt]{\includegraphics[width=0.14\linewidth]{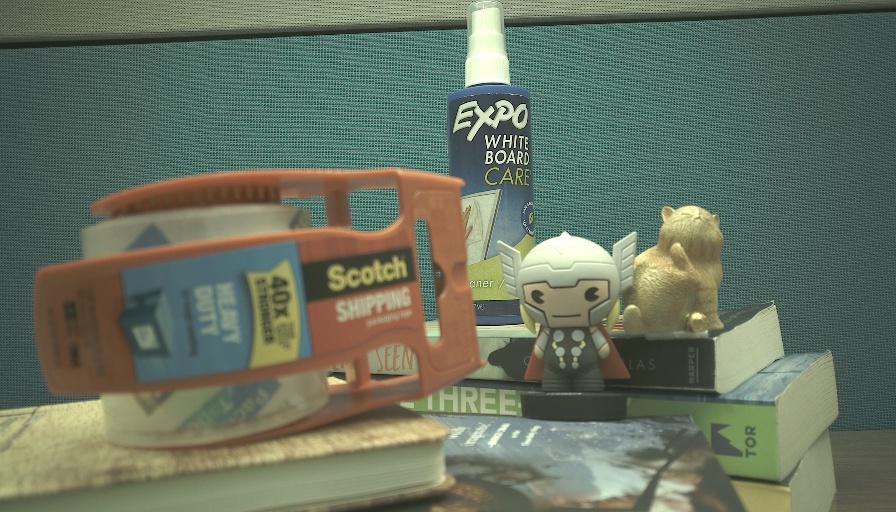}}{$I^1$}}	 &
    				\subfigure{
    					\stackunder[4pt]{\includegraphics[width=0.14\linewidth]{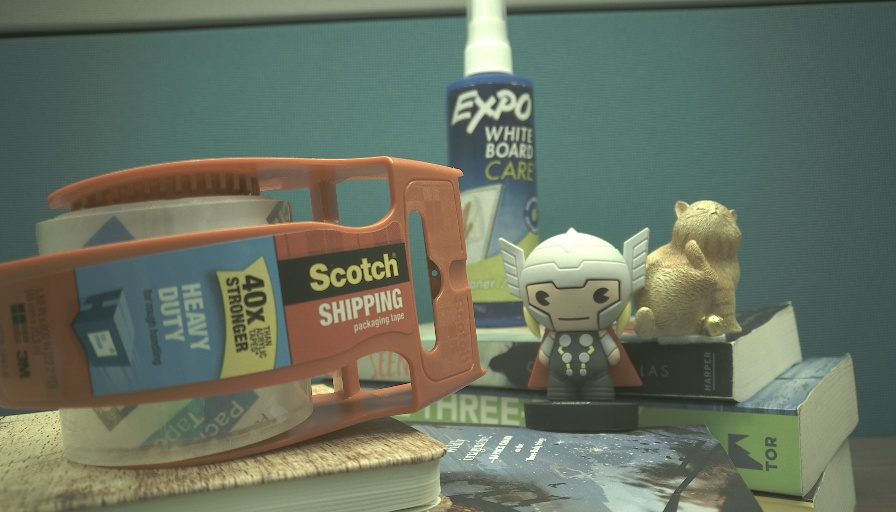}}{$I^2$}}	 &
    				\subfigure{
    					\stackunder[4pt]{\includegraphics[width=0.14\linewidth]{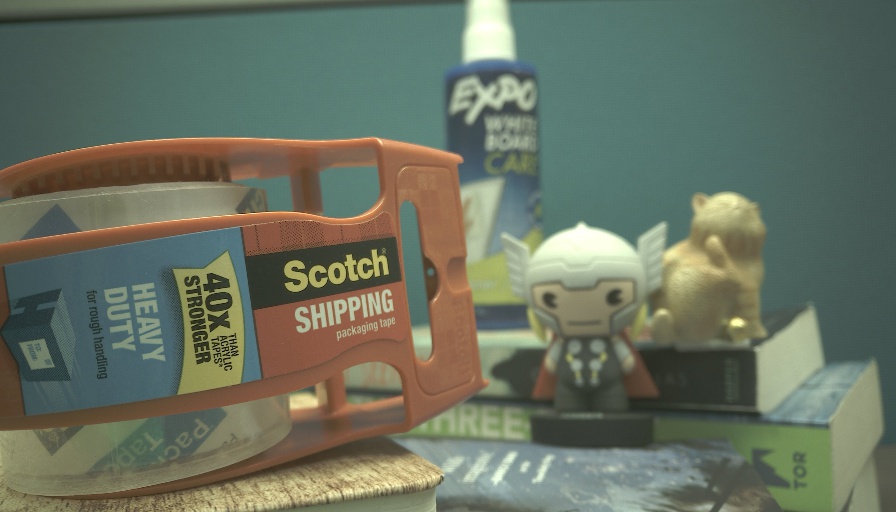}}{$I^3$}}	 &
    				\subfigure{
    					\stackunder[4pt]{\includegraphics[width=0.14\linewidth]{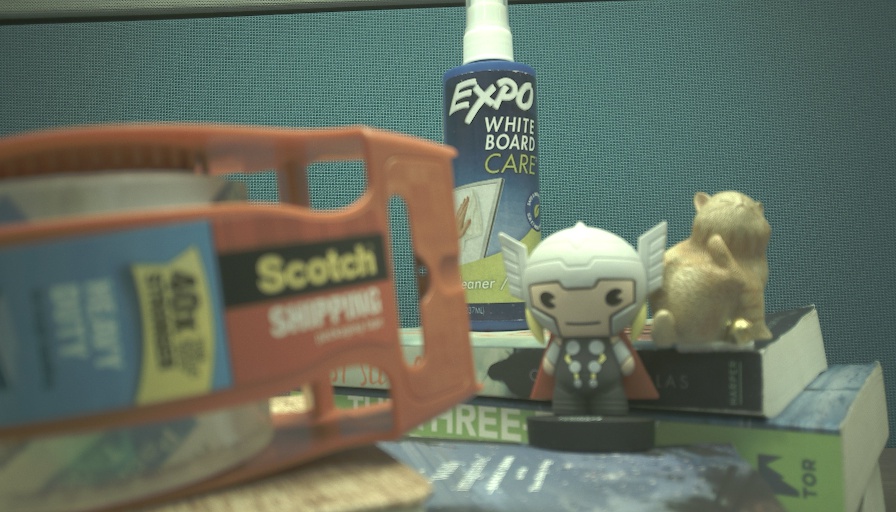}}{$I^4$}}	 &
    				\subfigure{
    					\stackunder[4pt]{\includegraphics[width=0.14\linewidth]{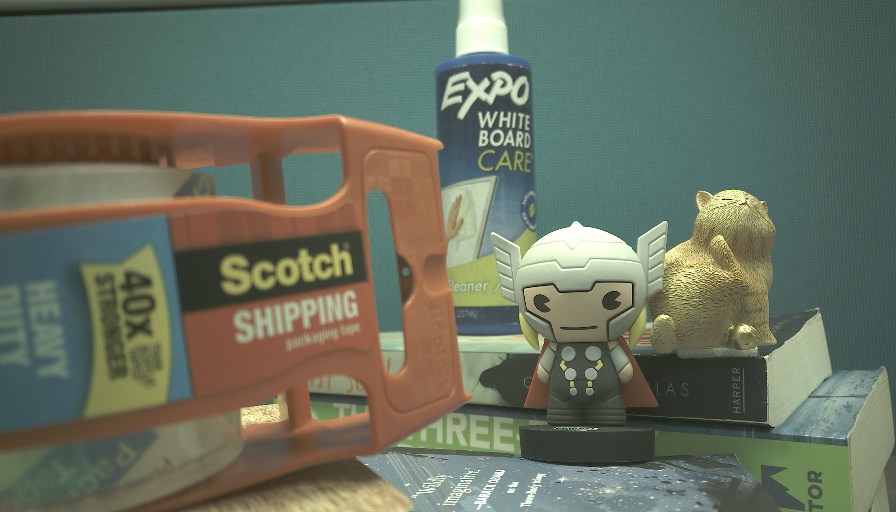}}{$I^5$}}	\\
    				\subfigure{
    					\stackunder[4pt]{\includegraphics[width=0.14\linewidth]{dfs_6frames_vertical/68302_s550_00.jpg}}{$J^0$}}		&
    				\subfigure{
    					\stackunder[4pt]{\includegraphics[width=0.14\linewidth]{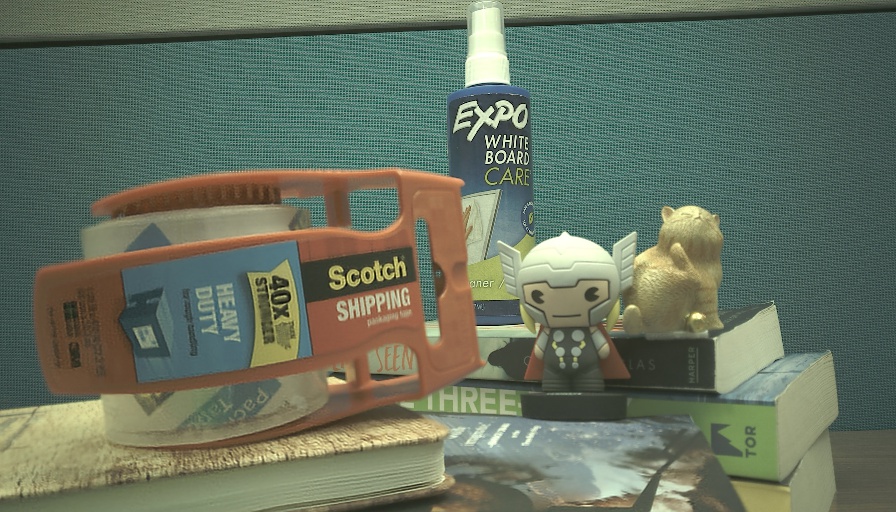}}{$J^1$}}	 &
    				\subfigure{
    					\stackunder[4pt]{\includegraphics[width=0.14\linewidth]{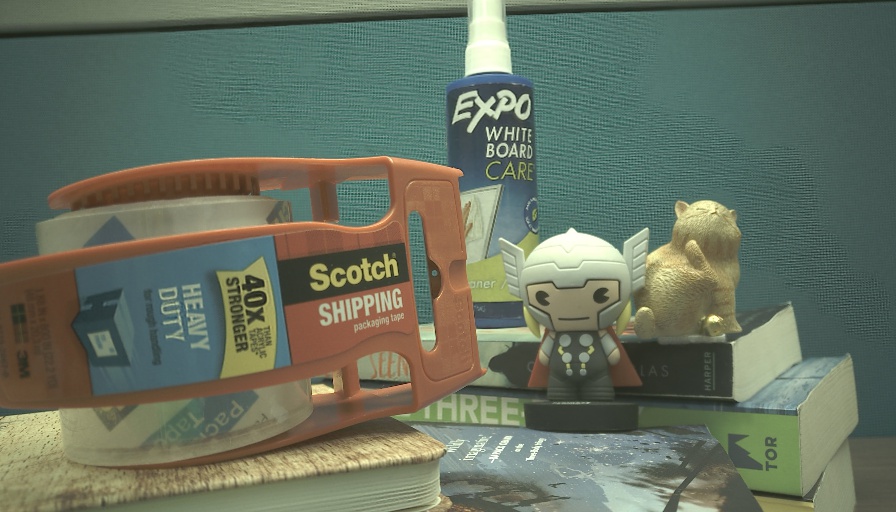}}{$J^2$}}	 &
    				\subfigure{
    					\stackunder[4pt]{\includegraphics[width=0.14\linewidth]{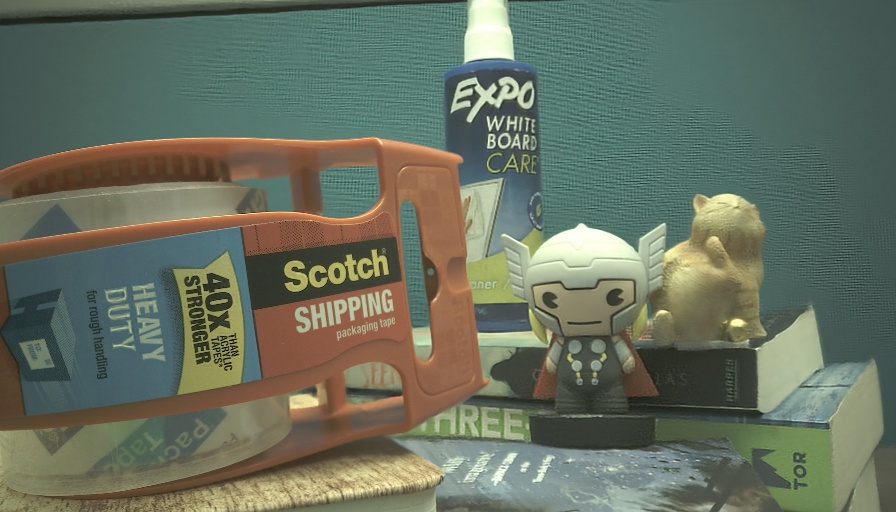}}{$J^3$}}	 &
    				\subfigure{
    					\stackunder[4pt]{\includegraphics[width=0.14\linewidth]{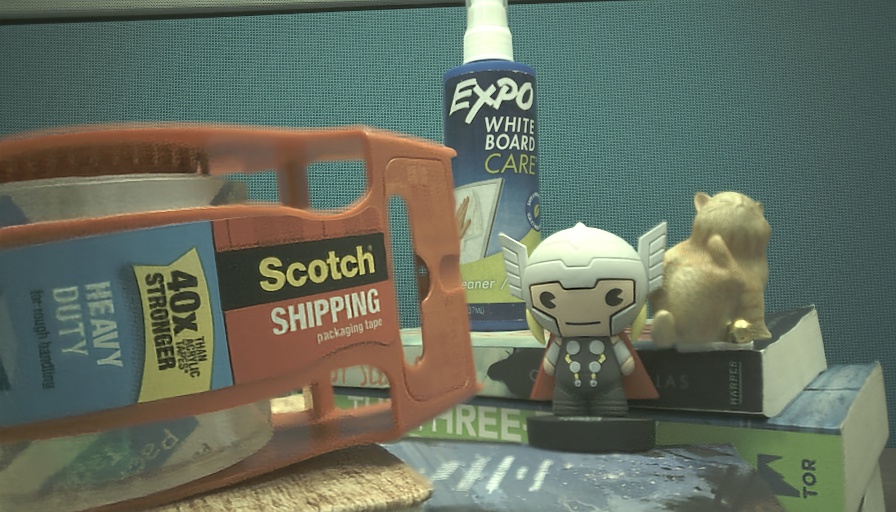}}{$J^4$}}	 &
    				\subfigure{
    					\stackunder[4pt]{\includegraphics[width=0.14\linewidth]{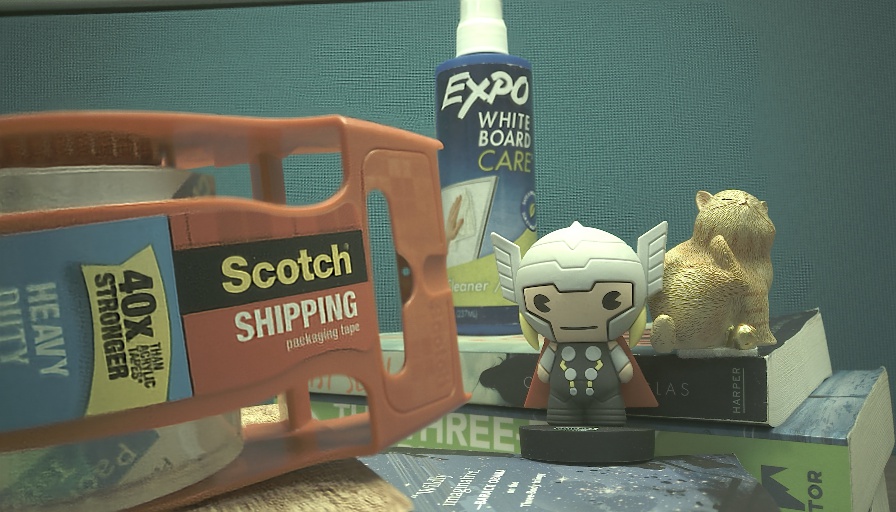}}{$J^5$}}
    			\end{tabular}
    		\end{minipage}
    \caption{Two all-in-focus frame sequences generated using the proposed focus strategy ($k=1$). $I^t$ and $J^t$ denote the captured frame and the fused frame at time step $t$. The camera made transverse movements in the first example and axial movements in the second example. Zoom to see the details.\label{fig: dynamic fs results}}
	}
\end{figure*}

Two sets of captured and fused frames are shown in Figure \ref{fig: dynamic fs results}, where $I$ denotes the captured frames and $J$ denotes the fused frames. In the two examples, the camera made transverse movements and axial movements respectively. The fused sequence demonstrate the effectiveness of the proposed strategy. 

\section{Conclusion}
\label{sec:conclusion}
Machine learning has brought a revolution in image processing and computer vision in the past decade. In contrast, data acquisition still relies primarily on traditional approaches. In this paper, we discuss a deep learning solution to a basic but not trivial control problem in image data acquisition: autofocus. We first propose an efficient deep learning autofocus pipeline. By introducing a step estimator and a focus discriminator, the proposed method achieves faster focus than traditional contrast maximization methods. We further extend the usage of the step estimator to all-in-focus imaging and show that it not only outperforms traditional focus stacking in static scene but also produces all-in-focus videos.



\bibliographystyle{IEEEtran}
\bibliography{references}

\begin{thebibliography}{10}
\providecommand{\url}[1]{#1}
\csname url@samestyle\endcsname
\providecommand{\newblock}{\relax}
\providecommand{\bibinfo}[2]{#2}
\providecommand{\BIBentrySTDinterwordspacing}{\spaceskip=0pt\relax}
\providecommand{\BIBentryALTinterwordstretchfactor}{4}
\providecommand{\BIBentryALTinterwordspacing}{\spaceskip=\fontdimen2\font plus
\BIBentryALTinterwordstretchfactor\fontdimen3\font minus
  \fontdimen4\font\relax}
\providecommand{\BIBforeignlanguage}[2]{{%
\expandafter\ifx\csname l@#1\endcsname\relax
\typeout{** WARNING: IEEEtran.bst: No hyphenation pattern has been}%
\typeout{** loaded for the language `#1'. Using the pattern for}%
\typeout{** the default language instead.}%
\else
\language=\csname l@#1\endcsname
\fi
#2}}
\providecommand{\BIBdecl}{\relax}
\BIBdecl

\bibitem{parallelCam}
D.~J. Brady, W.~Pang, H.~Li, Z.~Ma, Y.~Tao, and X.~Cao, ``Parallel cameras,''
  \emph{Optica}, vol.~5, no.~2, pp. 127--137, 2018.

\bibitem{yao2006evaluation}
Y.~Yao, B.~Abidi, N.~Doggaz, and M.~Abidi, ``Evaluation of sharpness measures
  and search algorithms for the auto focusing of high-magnification images,''
  in \emph{Visual Information Processing XV}, vol. 6246.\hskip 1em plus 0.5em
  minus 0.4em\relax International Society for Optics and Photonics, 2006, p.
  62460G.

\bibitem{kehtarnavaz2003development}
N.~Kehtarnavaz and H.-J. Oh, ``Development and real-time implementation of a
  rule-based auto-focus algorithm,'' \emph{Real-Time Imaging}, vol.~9, no.~3,
  pp. 197--203, 2003.

\bibitem{he2003modified}
J.~He, R.~Zhou, and Z.~Hong, ``Modified fast climbing search auto-focus
  algorithm with adaptive step size searching technique for digital camera,''
  \emph{IEEE transactions on Consumer Electronics}, vol.~49, no.~2, pp.
  257--262, 2003.

\bibitem{guo2018fast}
C.~Guo, Z.~Ma, X.~Guo, W.~Li, X.~Qi, and Q.~Zhao, ``Fast auto-focusing search
  algorithm for a high-speed and high-resolution camera based on the image
  histogram feature function,'' \emph{Applied Optics}, vol.~57, no.~34, pp.
  F44--F49, 2018.

\bibitem{gamadia2012filter}
M.~Gamadia and N.~Kehtarnavaz, ``A filter-switching auto-focus framework for
  consumer camera imaging systems,'' \emph{IEEE Transactions on Consumer
  Electronics}, vol.~58, no.~2, 2012.

\bibitem{yousefi2011new}
S.~Yousefi, M.~Rahman, and N.~Kehtarnavaz, ``A new auto-focus sharpness
  function for digital and smart-phone cameras,'' \emph{IEEE Transactions on
  Consumer Electronics}, vol.~57, no.~3, 2011.

\bibitem{gamadia2010performance}
M.~Gamadia and N.~Kehtarnavaz, ``Performance metrics for auto-focus in digital
  and cell-phone cameras,'' in \emph{Consumer Electronics (ICCE), 2010 Digest
  of Technical Papers International Conference on}.\hskip 1em plus 0.5em minus
  0.4em\relax IEEE, 2010, pp. 69--70.

\bibitem{park2008fast}
B.-K. Park, S.-S. Kim, D.-S. Chung, S.-D. Lee, and C.-Y. Kim, ``Fast and
  accurate auto focusing algorithm based on two defocused images using discrete
  cosine transform,'' in \emph{Digital Photography IV}, vol. 6817.\hskip 1em
  plus 0.5em minus 0.4em\relax International Society for Optics and Photonics,
  2008, p. 68170D.

\bibitem{chen2010passive}
C.-Y. Chen, R.-C. Hwang, and Y.-J. Chen, ``A passive auto-focus camera control
  system,'' \emph{Applied Soft Computing}, vol.~10, no.~1, pp. 296--303, 2010.

\bibitem{han2011novel}
J.-W. Han, J.-H. Kim, H.-T. Lee, and S.-J. Ko, ``A novel training based
  auto-focus for mobile-phone cameras,'' \emph{IEEE Transactions on Consumer
  Electronics}, vol.~57, no.~1, 2011.

\bibitem{mir2015autofocus}
H.~Mir, P.~Xu, R.~Chen, and P.~van Beek, ``An autofocus heuristic for digital
  cameras based on supervised machine learning,'' \emph{Journal of Heuristics},
  vol.~21, no.~5, pp. 599--616, 2015.

\bibitem{jiang2018transform}
S.~Jiang, J.~Liao, Z.~Bian, K.~Guo, Y.~Zhang, and G.~Zheng, ``Transform-and
  multi-domain deep learning for single-frame rapid autofocusing in whole slide
  imaging,'' \emph{Biomedical optics express}, vol.~9, no.~4, pp. 1601--1612,
  2018.

\bibitem{wei2018neural}
L.~Wei and E.~Roberts, ``Neural network control of focal position during
  time-lapse microscopy of cells,'' \emph{Scientific reports}, vol.~8, no.~1,
  p. 7313, 2018.

\bibitem{guojin2010image}
C.~Guojin, L.~Yongning, Z.~Miaofen, and W.~Wanqiang, ``The image auto-focusing
  method based on artificial neural networks,'' in \emph{Computational
  Intelligence for Measurement Systems and Applications (CIMSA), 2010 IEEE
  International Conference on}.\hskip 1em plus 0.5em minus 0.4em\relax IEEE,
  2010, pp. 138--141.

\bibitem{lee2008enhanced}
S.-Y. Lee, Y.~Kumar, J.-M. Cho, S.-W. Lee, and S.-W. Kim, ``Enhanced autofocus
  algorithm using robust focus measure and fuzzy reasoning,'' \emph{IEEE
  Transactions on Circuits and Systems for Video Technology}, vol.~18, no.~9,
  pp. 1237--1246, 2008.

\bibitem{rahman2008real}
M.~T. Rahman and N.~Kehtarnavaz, ``Real-time face-priority auto focus for
  digital and cell-phone cameras,'' \emph{IEEE Transactions on Consumer
  Electronics}, vol.~54, no.~4, 2008.

\bibitem{santos1997evaluation}
A.~Santos, C.~Ortiz~de Sol{\'o}rzano, J.~J. Vaquero, J.~Pena, N.~Malpica, and
  F.~Del~Pozo, ``Evaluation of autofocus functions in molecular cytogenetic
  analysis,'' \emph{Journal of microscopy}, vol. 188, no.~3, pp. 264--272,
  1997.

\bibitem{krotkov1988focusing}
E.~Krotkov, ``Focusing,'' \emph{International Journal of Computer Vision},
  vol.~1, no.~3, pp. 223--237, 1988.

\bibitem{yazdanfar2008simple}
S.~Yazdanfar, K.~B. Kenny, K.~Tasimi, A.~D. Corwin, E.~L. Dixon, and R.~J.
  Filkins, ``Simple and robust image-based autofocusing for digital
  microscopy,'' \emph{Optics express}, vol.~16, no.~12, pp. 8670--8677, 2008.

\bibitem{wu2012bilateral}
Z.~Wu, D.~Wang, and F.~Zhou, ``Bilateral prediction and intersection
  calculation autofocus method for automated microscopy,'' \emph{Journal of
  microscopy}, vol. 248, no.~3, pp. 271--280, 2012.

\bibitem{wang2018fast}
Y.~Wang, H.~Feng, Z.~Xu, Q.~Li, Y.~Chen, and M.~Cen, ``Fast auto-focus scheme
  based on optical defocus fitting model,'' \emph{Journal of Modern Optics},
  vol.~65, no.~7, pp. 858--868, 2018.

\bibitem{hovela1992learning}
L.~F. Hovela, ``Learning how to focus: adaptive control of an autofocus
  camera,'' in \emph{High-Resolution Sensors and Hybrid Systems}, vol.
  1656.\hskip 1em plus 0.5em minus 0.4em\relax International Society for Optics
  and Photonics, 1992, pp. 383--394.

\bibitem{krizhevsky2012imagenet}
A.~Krizhevsky, I.~Sutskever, and G.~E. Hinton, ``Imagenet classification with
  deep convolutional neural networks,'' in \emph{Advances in neural information
  processing systems}, 2012, pp. 1097--1105.

\bibitem{Kang_2014_CVPR}
L.~Kang, P.~Ye, Y.~Li, and D.~Doermann, ``Convolutional neural networks for
  no-reference image quality assessment,'' in \emph{The IEEE Conference on
  Computer Vision and Pattern Recognition (CVPR)}, June 2014.

\bibitem{zhou2014learning}
B.~Zhou, A.~Lapedriza, J.~Xiao, A.~Torralba, and A.~Oliva, ``Learning deep
  features for scene recognition using places database,'' in \emph{Advances in
  neural information processing systems}, 2014, pp. 487--495.

\bibitem{ren2018learning}
Z.~Ren, Z.~Xu, and E.~Y. Lam, ``Learning-based nonparametric autofocusing for
  digital holography,'' \emph{Optica}, vol.~5, no.~4, pp. 337--344, 2018.

\bibitem{liu2017multi}
Y.~Liu, X.~Chen, H.~Peng, and Z.~Wang, ``Multi-focus image fusion with a deep
  convolutional neural network,'' \emph{Information Fusion}, vol.~36, pp.
  191--207, 2017.

\bibitem{du2017image}
C.~Du and S.~Gao, ``Image segmentation-based multi-focus image fusion through
  multi-scale convolutional neural network,'' \emph{IEEE access}, vol.~5, pp.
  15\,750--15\,761, 2017.

\bibitem{tang2018pixel}
H.~Tang, B.~Xiao, W.~Li, and G.~Wang, ``Pixel convolutional neural network for
  multi-focus image fusion,'' \emph{Information Sciences}, vol. 433, pp.
  125--141, 2018.

\bibitem{guo2018fully}
X.~Guo, R.~Nie, J.~Cao, D.~Zhou, and W.~Qian, ``Fully convolutional
  network-based multifocus image fusion,'' \emph{Neural computation}, vol.~30,
  no.~7, pp. 1775--1800, 2018.

\bibitem{hasinoff2009time}
S.~W. Hasinoff, K.~N. Kutulakos, F.~Durand, and W.~T. Freeman,
  ``Time-constrained photography,'' in \emph{2009 IEEE 12th International
  Conference on Computer Vision}.\hskip 1em plus 0.5em minus 0.4em\relax IEEE,
  2009, pp. 333--340.

\bibitem{vaquero2011generalized}
D.~Vaquero, N.~Gelfand, M.~Tico, K.~Pulli, and M.~Turk, ``Generalized
  autofocus,'' in \emph{2011 IEEE Workshop on applications of computer vision
  (WACV)}.\hskip 1em plus 0.5em minus 0.4em\relax IEEE, 2011, pp. 511--518.

\bibitem{choi2017improved}
D.~Choi, A.~Pazylbekova, W.~Zhou, and P.~van Beek, ``Improved image selection
  for focus stacking in digital photography,'' in \emph{2017 IEEE International
  Conference on Image Processing (ICIP)}.\hskip 1em plus 0.5em minus
  0.4em\relax IEEE, 2017, pp. 2761--2765.

\bibitem{li2018scene}
W.~Li, G.~Wang, X.~Hu, and H.~Yang, ``Scene-adaptive image acquisition for
  focus stacking,'' in \emph{2018 25th IEEE International Conference on Image
  Processing (ICIP)}.\hskip 1em plus 0.5em minus 0.4em\relax IEEE, 2018, pp.
  1887--1891.

\bibitem{Agustsson_2017_CVPR_Workshops}
E.~Agustsson and R.~Timofte, ``Ntire 2017 challenge on single image
  super-resolution: Dataset and study,'' in \emph{The IEEE Conference on
  Computer Vision and Pattern Recognition (CVPR) Workshops}, July 2017.

\bibitem{clic}
\BIBentryALTinterwordspacing
``Workshop and challenge on learned image compression (clic),'' 2018. [Online].
  Available: \url{http://www.compression.cc/}
\BIBentrySTDinterwordspacing

\bibitem{tensorflow2015-whitepaper}
\BIBentryALTinterwordspacing
M.~Abadi, A.~Agarwal, P.~Barham, E.~Brevdo, Z.~Chen, C.~Citro, G.~S. Corrado,
  A.~Davis, J.~Dean, M.~Devin, S.~Ghemawat, I.~Goodfellow, A.~Harp, G.~Irving,
  M.~Isard, Y.~Jia, R.~Jozefowicz, L.~Kaiser, M.~Kudlur, J.~Levenberg,
  D.~Man\'{e}, R.~Monga, S.~Moore, D.~Murray, C.~Olah, M.~Schuster, J.~Shlens,
  B.~Steiner, I.~Sutskever, K.~Talwar, P.~Tucker, V.~Vanhoucke, V.~Vasudevan,
  F.~Vi\'{e}gas, O.~Vinyals, P.~Warden, M.~Wattenberg, M.~Wicke, Y.~Yu, and
  X.~Zheng, ``{TensorFlow}: Large-scale machine learning on heterogeneous
  systems,'' 2015, software available from tensorflow.org. [Online]. Available:
  \url{https://www.tensorflow.org/}
\BIBentrySTDinterwordspacing

\bibitem{kingma2014adam}
D.~P. Kingma and J.~Ba, ``Adam: A method for stochastic optimization,''
  \emph{arXiv preprint arXiv:1412.6980}, 2014.

\bibitem{QIU201935}
X.~Qiu, M.~Li, L.~Zhang, and X.~Yuan, ``Guided filter-based multi-focus image
  fusion through focus region detection,'' \emph{Signal Processing: Image
  Communication}, 2019.

\bibitem{bay2006surf}
H.~Bay, T.~Tuytelaars, and L.~Van~Gool, ``Surf: Speeded up robust features,''
  in \emph{Computer Vision -- ECCV 2006}, A.~Leonardis, H.~Bischof, and
  A.~Pinz, Eds.\hskip 1em plus 0.5em minus 0.4em\relax Berlin, Heidelberg:
  Springer Berlin Heidelberg, 2006, pp. 404--417.

\bibitem{cheng2019dual}
M.~Cheng, Z.~Ma, M.~S. Asif, Y.~Xu, H.~Liu, W.~Bao, and J.~Sun, ``A dual camera
  system for high spatiotemporal resolution video acquisition,'' \emph{arXiv
  preprint arXiv:1909.13051}, 2019.

\bibitem{Perazzi2016}
F.~Perazzi, J.~Pont-Tuset, B.~McWilliams, L.~{Van Gool}, M.~Gross, and
  A.~Sorkine-Hornung, ``A benchmark dataset and evaluation methodology for
  video object segmentation,'' in \emph{Computer Vision and Pattern
  Recognition}, 2016.

\bibitem{li2019learning}
Z.~Li, T.~Dekel, F.~Cole, R.~Tucker, N.~Snavely, C.~Liu, and W.~T. Freeman,
  ``Learning the depths of moving people by watching frozen people,'' in
  \emph{Proceedings of the IEEE Conference on Computer Vision and Pattern
  Recognition}, 2019, pp. 4521--4530.

\end{thebibliography}

\ifpeerreview \else






\fi

\end{document}